\documentclass[a4paper, 11pt, twoside]{report}
\pdfoutput=1

\usepackage[english]{babel}
\usepackage[utf8x]{inputenc}
\usepackage[T1]{fontenc}
\usepackage{float}
\usepackage{parskip}
\usepackage{textcomp}
\usepackage{gensymb}

\usepackage[a4paper,top=3cm,bottom=3cm,left=3cm,right=3cm,marginparwidth=2cm]{geometry}

\usepackage{amsmath}
\usepackage{graphicx}
\usepackage[colorinlistoftodos]{todonotes}
\usepackage[colorlinks=true, allcolors=blue]{hyperref}
\usepackage[numbers]{natbib}
\usepackage{url}
\usepackage{listings}
\usepackage{adjustbox}

\title{Issuing Green Bonds on the Algorand Blockchain}
\author{Gidon Katten}

\begin{document}
\begin{titlepage}

\newcommand{\HRule}{\rule{\linewidth}{0.5mm}} 


\includegraphics[width=8cm]{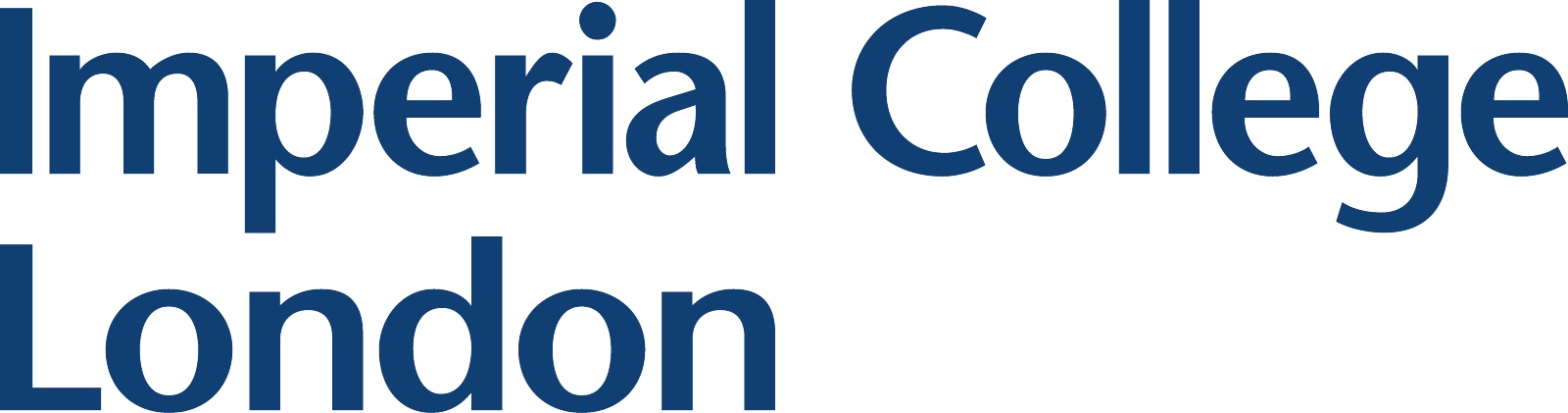}\\[1cm] 
 

\center 


\textsc{\LARGE BEng Individual Project}\\[1.5cm] 
\textsc{\Large Imperial College London}\\[0.5cm] 
\textsc{\large Department of Computing}\\[0.5cm] 

\makeatletter
\HRule \\[0.4cm]
{ \huge \bfseries \@title}\\[0.4cm] 
\HRule \\[1.5cm]
 

\begin{minipage}{0.4\textwidth}
\begin{flushleft} \large
\emph{Author:}\\
\@author 
\end{flushleft}
\end{minipage}
~
\begin{minipage}{0.4\textwidth}
\begin{flushright} \large
\emph{Supervisors:} \\
Prof. William Knottenbelt and Dr Maria Vigliotti \\[1.2em] 
\emph{Second Marker:} \\
Dr. Arthur Gervais 
\end{flushright}
\end{minipage}\\[2cm]
\makeatother



{\large \today}\\[2cm] 

\vfill 

\end{titlepage}

\begin{abstract}
Green bonds have been shown to be effective tool for sustainability however market growth is impeded by high issuance and transaction costs. The lack of appropriate standardisation and frameworks raise fear of greenwashing. 

In this paper, we propose a platform for green bond issuance on the Algorand blockchain. It offers \textit{Green Bonds as a Service}, increasing accessibility through automation. The solution has minimal associated costs and supports fractional asset ownership, both of which will help adoption especially in developing countries. A financial regulator must preapprove an investor and can freeze assets in the case of financial irregularities. Green bonds can be bought directly from an issuer or in the secondary market. We also introduce a novel mechanism whereby an issuer can upload proof of impact reports. A green verifier uses these to submit a green rating; poor green ratings result in reputational damage and economic penalties.

\end{abstract}

\renewcommand{\abstractname}{Acknowledgements}
\begin{abstract}
I would like to thank the following people:

\begin{itemize}
    \item My project supervisors Professor William Knottenbelt and Dr Maria Vigliotti for our insightful meetings and discussions throughout the project.
    \item Dr. Arthur Gervais for his advice early on in the project and his course, Principles of Distributed Ledgers.
    \item Dr. Enrico Biffis for his recommendations and ideas relating to the financial side of the project.
    \item My personal tutor, Dr. Krysia Broda, for her continued support throughout my time at Imperial College London.
\end{itemize}
\end{abstract}

\tableofcontents

\chapter{Introduction}

\section{Motivation}
Rapid change and innovation is needed across the world to ensure countries align themselves with the Paris Agreement of keeping the increase in global average temperature to well below 2\degree C. Green bonds are fixed income instruments used to fund environmentally friendly projects and can be a fundamental tool for climate investment. 

Currently the global green bond market represents approximately 2\% of the overall bond market \cite{sotm}. Growth is impaired by high issuance and transaction costs \cite{greenbondbarriers}, in particular for emerging markets where the size of issuance tends to be smaller \cite{greenbondemergingcountriesbarriers}. 

One has to look at the fundamentals of the green bond market to understand why a blockchain is an appropriate solution \cite{doyouneedablochchain}. With the exception of treasury bonds, investors must go through a broker to purchase bonds. A bond broker acts as a intermediary between buyers and sellers and may take advantage of the lack of price transparency by marking up prices. In addition, each trade typically has to go through a settlement period, where one has to wait a few days between the trade agreement and the date at which the trade is considered final.

Specific to green bonds, there is an added relationship between the issuer and investor. An investor wants to avoid contributing to \textit{greenwashing}, whereby the funds generated from the bond are not actually used for green projects. The issuer must produce data and outline the use of proceeds to satisfy investor distrust. At this time, there are no mechanisms to enforce issuer compliance \cite{issuercompliance}.

Blockchain presents a number of opportunities. Blockchain can help streamline the clearing and settlement process of green bonds through disintermediation \cite{esma}. Transparent smart contracts would automatically perform atomic transactions, reducing costs and counterparty risk. Blockchain also enables self-custody whereby issuers and investors alike can securely store and manage their assets. This not only eliminates custodian fees, but also helps democratise the green bond market as one does not need access to a financial intermediary service.

Green bond issuance, irrespective of an issuers size, would face similar fees, opening up the retail market. Blockchain can also help automate green bond reporting and provide a structure for green accountability. A credible green bond market can be built by integrating use of proceeds and proof of impact reporting.

Blockchain has been receiving negative media attention regarding its energy consumption. The University of Cambridge Bitcoin Consumption Index \cite{cbeci} estimates Bitcoin uses similar amounts of energy per year compared to countries such as Argentina and the Netherlands. Some would argue this subsequently makes blockchain a bad candidate for green bond issuance, where green investors may be reluctant to use such a technology.

Algorand uses a Proof of Stake (PoS) consensus mechanism whereas Bitcoin and Ethereum use Proof of Work (PoW). PoS consumes several orders of magnitude less energy than PoW \cite{blockchainenergyconsumption}. Ethereum is working on switching to PoS but it remains to be seen when this will happen. This makes Algorand an ideal candidate for green bond issuance. Furthermore the Algorand team claim their blockchain is carbon neutral \cite{algorandcarbonneutral}. Whilst there has yet to be independent research on Algorand's energy consumption, many environmentally focused companies are leveraging Algorand \cite{planetwatch}.

The Algorand blockchain is a permissions blockchain so everyone can review the market and make informed investment decisions. It has a high transaction throughput and low transaction costs, making it accessible to smaller investors. Algorand also has native support for smart contracts and tokens which benefit from the same speed and cost as any other transaction. 


\section{Objectives}
\label{section:objectives}
The objectives for the project were:
\begin{enumerate}
    \item Developing a solution to issue green bonds on the blockchain and assess benefits for all stakeholders (issuer, investor, green verifier and financial regulator).
    \item Investigate the Algorand blockchain and assess how its unique features can benefit the project.
\end{enumerate}

As part of the first objective, it was important to develop a platform, as a proof of concept, that would support the green bond life cycle. This includes issuance, coupons, principal, financial regulation and green verification. 

Algorand is a relatively new technology and does not yet have a mature set of developer tools. Therefore the project serves the purpose of discovering if Algorand is ready to support such a product.

\section{Contributions}
The main contributions of the project are:

\begin{itemize}
    \item \textbf{Deploying the bond structure on the blockchain} An issuer can specify the bond parameters and investors can receive coupons, and the principal at maturity. 
    \item \textbf{Primary and secondary market} Investors can purchase green bond directly from an issuer or trade them among themselves in the secondary market. The exchange of the bond and payment are implemented using atomically grouped transactions so if one fails then they both fail. The exchange is confirmed in less than five seconds.
    \item \textbf{Fractional asset ownership} An investor can buy and sell parts of a green bond. This opens up the market to smaller investors which may not have the funds to purchase a whole bond or would like to recoup some of their money.
    \item \textbf{Green verification and rating} The issuer can upload reports regarding the green status of the bond which the green verifier can read when determining the appropriate green rating to give. Bad ratings negatively impact the coupon payments the issuer pays which should help address greenwashing.
    \item \textbf{Financial regulation} All issuers / investors must be preapproved by the financial regulator before they can issue / invest in a green bond. The financial regulator also has the ability to freeze the green bond and prevent any associated payments whether for everyone or specified accounts. This functionality can be used to fulfil KYC and AML obligations.
    \item \textbf{Decentralized application} A web application where users can issue green bonds on the Algorand blockchain. One can take the role of an issuer, investor, green verifier or financial regulator and see how they interact with one another.
\end{itemize}

\chapter{Background}
\label{chapter:background}
In this chapter, we will introduce green bonds and their regulatory frameworks. We will review the current state of the global green bond market, relating to both traditional and blockchain issuance. We will discuss the key concepts surrounding blockchain, some examples of blockchains and relevant considerations when selecting a blockchain to use. We will end by covering the Algorand blockchain in more detail. 


\section{Bonds}
\label{section:bonds}
A bond is an agreement to pay money according to the rules specified at issuance. The bond parameters include:
\begin{itemize}
    \item \textbf{Maturity Date} The date of the last payment.
    \item \textbf{Coupon} An amount paid periodically up until maturity. 
    \item \textbf{Principal} An amount paid at maturity. Also known as face value or par value.
\end{itemize}

\begin{figure}[H]
\includegraphics[scale=0.9]{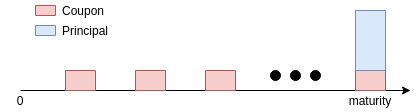}
\centering
\caption{Bond Payout Structure}
\label{fig:bondpayout}
\end{figure}

Figure \ref{fig:bondpayout} shows the timeline, between issuance and maturity, for when the coupons and principal are payed out. Variations of bonds exist, for example a zero-coupon bond is a bond with no coupons. The coupon is typically expressed using the coupon rate, which is equal to the coupon amount divided by the principal amount. Let's say we have a bond that matures after 5 years with a face value of \$100 and a coupon rate of 10\% payed annually. At the end of each subsequent year, it pays \$10 for the coupon and at maturity it pays \$100 on top of the coupon.

Assuming a coupon payment has just been made, the theoretical price of a bond can be determined using the following formula:

\begin{align*}
P &= \sum_{t=1}^{T} \frac{C}{\left(1+r\right)^t} + \frac{F}{\left(1+r\right)^T} \\
&= C \left[ \frac{1-(1+r)^{-T}}{r}) \right] + \frac{F}{\left(1+r\right)^T} \\
\end{align*}
where $C$ is the coupon payment at each time period, $r$ is the discount rate (yield to maturity), $F$ is the face value of the bond and $T$ is the number of time periods until maturity.

\section{Green Bonds}
Green bonds are structured in the same way as regular bonds. An issuer creates the bond to raise funds, and investors receive coupons and a principle when it matures. Like the name suggests, the proceeds of the bond must be used to benefit the environment and are therefore subject to additional checks and regulations. Typically, a second or third party approves the use of proceeds from the bond and all related expenditures. Often there is also ongoing reporting on the status of the project to ensure the green key performance indicators are being met.

It may seem unclear why companies would issue green bonds. They inherit the properties of conventional bonds but have the disadvantage of extra restrictions as well as expensive procedures needed to prove compliance. However green bonds do have several advantages over their generic counterparts. 

Green bonds has been shown to attract impact investors and increase institutional ownership \cite{shares}, with a stronger response for first time issues and bonds certified by third parties. Investors may also accept lower returns as they prioritise the environment over financial gain \cite{premium}. 

Green bonds are also an effective tool for sustainability which is reflected in higher share prices as companies credibly signal their commitment to the environment \cite{cgb}. Following the issuance of green bonds, we observe improvements in a company's environmental performance, in terms of lower CO\textsubscript{2} emissions and higher environmental ratings \cite{shares}.

\subsection{Green Bond Standards}
\label{subsection:standards}
There are no universal standards when it comes to green bonds. Two frameworks commonly used are the Green Bond Principles (GBP) \cite{gbp} created by the International Capital Market Association (ICMA) and the Climate Bond Standard (CBS) \cite{cbs}, developed by the Climate Bonds Initiative (CBI).

The GBP are a set of voluntary guidelines that can be broken down to four core components:
\begin{enumerate}
    \item Use of Proceeds: Outlines the eligible categories for green projects. These range from renewable energy to sustainable water and wastewater management.
    \item Process for Project Evaluation and Selection: Encourages transparency regarding how the project falls within the eligible categories and any potential environmental and social risks associated with the project. Recommends an external review to supplement the review.
    \item Management of Proceeds: Should track funds and inform investors the placement of unallocated net proceeds. Recommends the use of a third party to verify the internal management of proceeds.
    \item Reporting: Should continually update information surrounding the use of proceeds until full allocation. The report should be annual and in cases of material developments.
\end{enumerate}

The voluntary nature of GBP raises concerns of greenwashing, which has led to the development of CBS, a certification scheme based on the underlying principles of GBP. The certification can be separated into two sets of requirements: pre-issuance and post-issuance. For pre-issuance, there are mandatory requirements for the use of proceeds which must be independently verified by an approved body. For post-issuance, the issuer must annually report on the allocation and eligibility for the project however proof of impact is optional.

According to the Climate Bonds Initiative \cite{sotm}, 86\% of green bonds issued in 2019 have some form of external review, of which almost two-thirds used second party opinions and 17\% used the CBS certification. The adoption levels can be partially explained by the cost of certification, which is estimated between USD 10 and 100 thousand \cite{greenbondbarriers}.

\subsection{Green Bond Market}
The history of the green bond market begins in 2007 when the European Investment Bank issued the first green bond, then called \textit{Climate Awareness Bond}. The bond proceeds were targeted to renewable energy and energy efficient projects. The following year, the World Bank issued the green bond as we know today. CICERO was used as second opinion provider to certify the \textit{greenness} of the bond and the World Bank incorporated impact reporting. More recently in the UK, the Chancellor Rishi Sunak announced Britain will issue its first green government bond in 2021. 

\begin{figure}[H]
\includegraphics[scale=0.4]{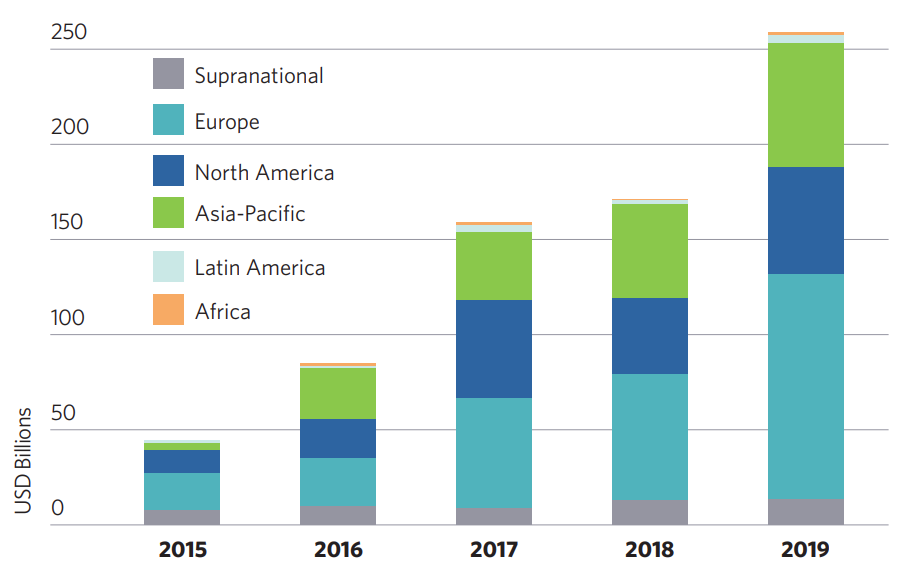}
\centering
\caption{Green bond issuance by region \cite{sotm}}
\label{fig:market}
\end{figure}

At that time, 2019 became the largest year for the global green bond market \cite{sotm}. USD 258.9 billion worth of bonds was issued, bringing the cumulative issuance since 2007 to USD 754 billion. Fig \ref{fig:market} shows the market size is growing, especially in Europe. Overall green bonds represent between 2-3\% of the total bond market share, leaving room for significant expansion.


\subsection{Blockchain Green Bonds}
To date there has only been a handful of blockchain green bond, highlighting the need for further research in this area. In early 2019, BBVA Group became the first to issue a blockchain green bond. The bond was labelled green by DNV GL, a second party opinion provider. A permissioned blockchain (see section \ref{subsection:PoWVsPoS}) was used to issue the bond, and a copy of the transaction was recorded in the permissionless Ethereum Testnet blockchain. This allowed BBVA to meet its Know Your Customer (KYC) requirements whilst providing transparency to the public.

A report by HSBC and Sustainable Digital Finance Alliance outlined three areas where blockchain can be applied to green bonds \cite{hsbc}:
\begin{enumerate}
    \item Structuring, issuance and distribution
    \item Transfer of ownership, payment and settlement
    \item Benchmarking and reporting
\end{enumerate}

The last area is specific to green bonds, whereas the first two also apply to conventional bonds. We can therefore use the more explored regular blockchain bonds as a reference to how the technology could be used for green bonds. 

In September 2019, Santander issued the first end-to-end blockchain bond \cite{santandercasestudy}. They used the public Ethereum blockchain however the bonds were only available to internal group companies. They both tokenized the USD 20 million debt and settled it with ERC-20 tokens. A smart contract was used ensure ownership eligibility as the Ethereum blockchain is permissionless.

\section{Blockchain}
\begin{figure}[H]
\includegraphics[scale=0.55]{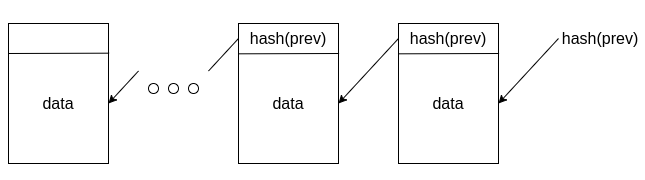}
\centering
\caption{Blockchain: A linked list using hash pointers}
\label{fig:blockchain}
\end{figure}

A blockchain is a distributed ledger in which blocks are linked in a chain using hash pointers. It is the underlying technology of Bitcoin and other cryptocurrencies. The technology itself provides a secure mechanism for storing ordered data, and eliminates the need for a central authority. Instead of pointing to the location of a block, a hash pointer points to a block's data. These pointers refer back to the previous block in the chain such that there is a single genesis block which is the root and does not point to anything. Fig \ref{fig:blockchain} shows the structure of a blockchain.

A collision resistant cryptographic hash function is used to ensure it is infeasible to create two blocks with the same hash. If someone were to tamper with a block then its hash would change. The next block in the sequence's hash pointer would no longer match the hash of the block it is pointing to which would invalidate the chain. Therefore an adversary would also have to tamper the next blocks hash, and so on, all the way to the head of the list. Therefore a given block in the chain guarantees the integrity of the whole history of the blockchain up to that point. As long as the head of the blockchain is agreed upon, and the cryptographic hash function is collision resistant, then the blockchain is tamper proof.

There are a variety of approaches to processing, authenticating and validating blocks. Permissioned blockchains, where only trusted actors can participate in reading and writing, can operate with much lower levels of protection. The disadvantage of this type of blockchain though is its closeness and centralisation, and for this reason, will not be the focus of this paper. Permissionless blockchains are open to the public and are fully transparent. Anyone can participate and read the on-chain data, which unlocks the green bond market to a wider investor population. Due to their openness, permissionless blockchains require extra protection to prevent adversarial attacks on the integrity of the blockchain.

\subsection{Proof of Work vs. Proof of Stake}
\label{subsection:PoWVsPoS}
There are a number of ways of achieving consensus in permissionless blockchains. Two of these are Proof of Work (PoW) and Proof of Stake (PoS). 

In PoW, miners race to solve a difficult cryptographic puzzle and the first one able to do so is rewarded with a cryptocurrency payment. The most famous example of a cryptocurrency that uses PoW is Bitcoin. Included in the block header is an arbitrary number called a nonce. Miners change the nonce till they found one such that the hash of the block is less than or equal to the current target hash value set by the network. The puzzle is designed in such a way that once a solution has been found, it is easy for the other nodes to verify that it is indeed a correct solution and only then is it used to append a block to the end of the blockchain.

The security of PoW relies on the difficulty of these puzzles. To achieve a fork in the chain, one would have to mine blocks faster than everyone else in the network. This would require large amounts of computing power which would be expensive and suffer from a significant opportunity cost. Even then, a fork may not succeed if the community decides not accept the longer chain.

One problem with PoW is the energy inefficiency \cite{energy}. Green bond investors may be reluctant to use technologies that have such cost even if it may be fractional compared to the effect of the bond. For this reason, this paper will focus on blockchains that use PoS where little energy is needed to mine blocks.

In PoS, the influence each node has is proportional to the number of tokens it has. A subset of nodes are chosen at random and together create and validate a proposed block. Algorand, the blockchain of choice for this project, achieves consensus through a pure PoS protocol \cite{algorandPPoS} based upon a decentralized Byzantine Agreement protocol. Section \ref{section:algorand} covers Algorand in more detail.

With PoS, the users with the most influence also are the most invested in the cryptocurrency. This serves as a strong incentive for these users to be honest because they have the most to lose if trust is lost in the technology and the currency depreciates. In addition, PoS removes the need for wasteful energy consumption making it appropriate for green bond issuance.

Another criticism leveraged at PoW that PoS addresses is its barrier to entry for mining. In PoW, miners often compete with each other using specialist hardware. This can lead to centralisation as a few large mining centres control the market \cite{pooledmining}. Comparatively, PoS enables more nodes to join the network and participate in the consensus protocol.

One potential issue with PoS is known as the \textit{nothing at stake} problem. In PoW mining, an attacker would have to consume a lot of resources attempting to create a fork in the chain. With naive PoS, at no extra cost one can continually try to create a fork while simultaneously mining on top of current longest chain. To solve this issue, PoS blockchains utilise mechanisms that punish nodes that are building blocks on more than one chain.

\subsection{Smart Contracts}

A smart contract is a computer program written to a blockchain that can be used to automatically enforce contracts. Smart contracts facilitate the transfer of digital assets between parties once predetermined conditions are met, without the need for a central authority. The code of the smart contract determines its functionality and often cannot be altered once added to the blockchain.

Smart contracts have many interesting applications \cite{smartContract}. Specific to green bonds, they can facilitate issuance and settlements, as well as automate investors coupons and principal payments. This presents opportunities to standardise the green bond offerings and provide \textit{bonds as a service} \cite{hsbc} \cite{sustainableInvestmentBlockchain}, where people can create their own green bonds at low costs.

An important property of smart contracts is its openness. Since they are written to a blockchain, their contents can be viewed by everyone. Trust is very important in the bond market so smart contracts present an opportunity to increase trust through its transparency \cite{transparency}. 


\subsection{Decentralised Applications}
Decentralised Applications (DApps) are applications with their backend code running on a decentralised peer-to-peer network. They typically have a frontend user interface and the backend utilises smart contract functionality. These applications benefit from its decentralised properties, making them more trustworthy and secure.

\subsection{Tokens and Stablecoins}
\label{subsection:stablecoin}
Tokens are assets that can be transferred using blockchain technology. Tokens can be fungible or non-fungible and can represent anything from a security to a voting right. Bitcoin is one such example of a fungible token. Tokens should not be used to store large amounts of data because storing data on a blockchain is very expensive. Instead this data should be stored externally, and its hash which acts as a fingerprint can be stored on the blockchain and tracked by tokens. Through this approach, tokens can be used for green reporting.

Another application of tokens is stablecoins, which address the problem of cryptocurrency volatility. Bond issuers and investors are unlikely to want to expose themselves to the risk of rapid price changes. Stablecoins can be divided into a number of categories depending on what, if anything, it is collateralized against. These are: fiat-collateralized, commodity-collateralized, crypto-collateralized and non-collateralized. The most common type and the type which provides the greatest stability, fiat-collateralized stablecoins, are digital assets that track fiat currency using smart contracts. 

A key difference between fiat-collateralized stablecoins and their corresponding fiat currency is where there risk fall. Fiat currency is guaranteed by the Central Bank however a stablecoin is not. One well known stablecoin is Tether, which is tied to the US Dollar. Tether originally claimed to have one-to-one reserves but have since admitted to using loans from affiliated entities. Using Tether and similar stablecoins requires trusting them and the commercial banks that are holding their reserves. Table \ref{tab:stablecoins} summarises the different types of fiat-collateralized stablecoins and their respective risks.

\begingroup
\renewcommand{\arraystretch}{1.5}
\begin{table}[H]
    \centering
    \begin{tabular}{c|c|c}
        \textbf{Issuer} & \textbf{Examples}                                                                              & \textbf{Risks}                                                                                          \\ \hline
        Central Bank    & There are currently no examples                                                                & National currency                                                                                       \\ \hline
        Bank            & JPMCoin                                                                                        & \begin{tabular}[c]{@{}c@{}}Bank holding the reserves\\ National currency\end{tabular}                   \\ \hline
        Consortium Bank & Fnality                                                                                        & \begin{tabular}[c]{@{}c@{}}Risk spread across the banks \\ National currency\end{tabular}               \\ \hline
        Finance Company & \begin{tabular}[c]{@{}c@{}}Tether (USDT)\\ USD Coin (USDC)\end{tabular} & \begin{tabular}[c]{@{}c@{}}Finance company\\ Bank holding the reserves\\ National currency\end{tabular}
    \end{tabular}
    \caption{Fiat-collateralized stablecoin types}
    \label{tab:stablecoins}
\end{table}
\endgroup



\section{Algorand}
\label{section:algorand}
Algorand \cite{algorand}, founded in 2017 by Turing Award winner Silvio Micali, is an open source public blockchain that uses proof of stake to achieve consensus. The currency used in Algorand is \textit{Algos}, and all online users with \textit{Algos} are eligible to participate in selecting and writing a given block into the blockchain.

The Algorand blockchain is also non-forking. In a given round, at most one block is certified and written to the chain. We can use its property of finality to confirm transactions in under five seconds.

When talking about blockchain, it is common to refer to layer-1 and layer-2. Layer-1 refers to the underlying main blockchain architecture and layer-2 the overlaying network that is built on top of the blockchain. Algorand anticipated many of its use cases and provided layer-1 features to solve them. In the subsequent sections, we will explain some of the features utilised in this project.

\subsection{Accounts}
An account in Algorand is made up of a (unique) public address and a corresponding private key. Each account has its own associated on-chain data, for example its Algo balance. 

A person can sign a transaction from their account using its' private key and submit it to the blockchain. There are various types of transactions that are supported, the most relevant of which are:
\begin{itemize}
    \item \textbf{Payment} Transfer Algos to an account
    \item \textbf{AssetTransfer} Send ASAs (see section \ref{subsection:asa}) to an account
    \item \textbf{ApplicationCall} Call a stateful smart contract (see section \ref{subsection:statefulsmartcontracts})
\end{itemize}

\subsection{Algorand Standard Assets}
\label{subsection:asa}
Algorand Standard Assets (ASAs) are assets built into layer-1 of the Algorand blockchain. They share the same properties as Algos in that they can be held by different addresses and can be transferred between parties.

ASAs can represent stablecoins, digital art, securities etc. When creating an asset you specify its parameters like the total number you would like to mint. Each created ASA is given a unique identification number. Just like with any currency, you cannot distinguish between instances of the same ASA.

To receive an ASA, one must first opt into it. This is because otherwise someone could send an account many meaningless assets which would pollute their holdings and incur a fee (see section \ref{subsection:fees}).

ASAs also offer additional functionality, two of which we will make use of. The first is that the asset creator can specify an account which has the ability to freeze all assets or freeze asset holdings for a specific account. The second is that the asset creator can set a clawback account that is allowed to revoke assets from one account and send these to another account. We will later combine these features in our implementation, section \ref{subsection:greenbond}.

\subsection{Atomic Transfers}
\label{subsection:atomictransfers}
Trading money and assets traditionally requires a trusted intermediary which ensures that both parties receive what they expect. Algorand solves this problem with its layer-1 Atomic Transfer protocol. 

In Algorand, you can group multiple transactions and submit them at one time. If any of the transactions within the group fail then they all do, and if they all succeed then the transactions are successful. A maximum of 16 transactions can be grouped together and these transactions can be of all different types. For example you can group a \textit{Payment} transaction and an \textit{AssetTransfer} transaction to exchange Algos for an asset. 

\subsection{Smart Contracts}
\label{subsection:smartcontracts}
Smart contracts in Algorand can be split into two categories: stateless and stateful. They are both written in the Transaction Execution Approval Language (TEAL), see section \ref{subsection:teal}. Often stateful and stateless contracts are linked together. This is done when you need to store some persistent data (stateful), whether global or local, which is paired with some type of spending transaction (stateless).

\subsection{Stateless Smart Contracts}
\label{subsection:statelesssmartcontracts}
Stateless smart contracts can be further divided into two models of use: contract accounts and delegated signatures. In both cases the contracts are not stored on the blockchain and their logic is evaluated at the time when a transaction is submitted. 

Contract accounts are like any other Algorand account, they have an address, can send and receive Algos and ASAs and can call stateful smart contracts. The only difference is that instead of using a private key to approve transactions from itself, it uses its associated stateless smart contract. This is known as a \textit{Logic Signature}. 

\begin{figure}[H]
\includegraphics[scale=0.8]{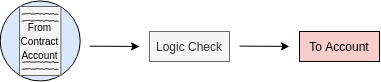}
\centering
\caption{Transaction from contract account}
\label{fig:contractaccount}
\end{figure}

Contract accounts can therefore be used as an escrow account. The address may own some Algos and ASAs and under certain conditions determined by its logic, one can transfer some of these holdings to another account. Fig \ref{fig:contractaccount} shows the process of sending a transaction from a contract account.

\begin{figure}[H]
\includegraphics[scale=0.72]{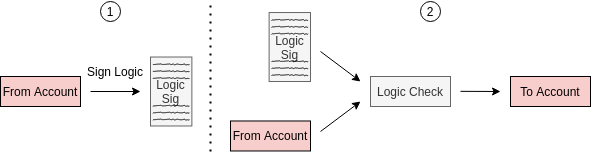}
\centering
\caption{Signing TEAL logic and submitting transaction from account using delegated signature}
\label{fig:delegatedsignature}
\end{figure}

Delegated signatures is another way of using stateless smart contracts. An Algorand account can sign a stateless smart contract with their private key. At this point then someone can use this stateless smart contract to sign transactions on behalf of the original account that signed the logic. This two stage process is outlined in Fig \ref{fig:delegatedsignature}.

An example use case for delegated signatures would be someone signing logic which can be used by their gas company to extract payment from them. The stateless smart contract can be written and signed to approve a monthly transfer of 100 Algos from their account to the gas company.

\subsection{Stateful Smart Contracts}
\label{subsection:statefulsmartcontracts}
\begin{figure}[H]
\includegraphics[scale=0.7]{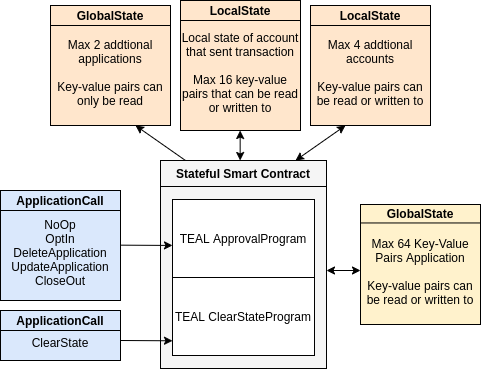}
\centering
\caption{Algorand Stateful Smart Contract}
\label{fig:stateful}
\end{figure}

Stateful smart contracts, also known as \textit{applications},  live on the blockchain and their logic is evaluated during block assembly time. Stateful smart contracts differ with stateless smart contracts in that they do not approve spending transactions from an account, but rather they approve logic that manipulates its stored state. If the application call transaction returns false, any changes to its state triggered by the transaction will be cancelled. Fig \ref{fig:stateful} shows an overview of the stateful smart contract in Algorand.

Stateful smart contracts utilise on chain storage by storing additional values both in a global state and a local state. An application has a single global state which is accessible to any account that calls the smart contract. To access local state, one must first opt into the application. Each opted in account has its own local state.

The global state of the contract is limited to 64 key-value pairs, and the local state is limited to 16 key-value pairs for each individual account that interacts with it. The cost of the contract is proportional to the amount of storage used. The creator account must fund the global state cost while each opted in account is responsible for its own local storage cost, see section \ref{subsection:fees} for more details.

Stateful smart contracts are made up of two programs: the \textit{ApprovalProgram} and the \textit{ClearStateProgram}. There are a set number of supported application calls; the \textit{ClearState} transaction type is handled by the \textit{ClearStateProgram} and all others are handled by the \textit{ApprovalProgram}:

\begin{itemize}
    \item \textbf{OptIn} Enable local state storage
    \item \textbf{NoOp} Generic application call
    \item \textbf{UpdateApplication} Update the TEAL program with new logic code 
    \item \textbf{DeleteApplication} Delete the application
    \item \textbf{CloseOut} Opt out of application and its TEAL logic
    \item \textbf{ClearState} Like CloseOut but will always opt out of the application whether the transaction succeeds or fails
\end{itemize}

A transaction can include arguments as well as the IDs of additional applications and accounts. The global state of the applications referenced can only be read, and the local state of caller and referenced accounts can both be read and written to. In addition, the local state of the five mentioned accounts can be read for any smart contract written to the blockchain.

As seen above in the supported application calls list, Algorand has a unique functionality that allows you to update and delete a stateful smart contract. By default anyone can perform these actions so additional logic must be added to the \textit{ApprovalProgram} to prevent this. State requirements for the contract must be specified at creation and is the one exception in that it cannot be altered once the contract has been written to the blockchain.

\subsection{TEAL}
\label{subsection:teal}
Algorand smart contracts are written in an assembly like language called Transaction Execution Approval Language (TEAL) and is processed with a stack machine. TEAL contains a set of operators that interact with the values on the stack, for example \textit{pop} and \textit{bz}. There is scratch space which allows you to store and retrieve values in a program. TEAL also supports passing byte array arguments to a program.

Smart contracts can also be written in Python and compiled to TEAL using PyTeal. PyTeal introduces some basic control flow operations such as \textit{If} and \textit{Cond} but still remains very primitive. The majority of its supported statements can be directly mapped one-to-one with its equivalent in TEAL making PyTeal a simple wrapper. The compiler for PyTeal to TEAL performs no optimisations as well so one has to ensure they write efficient code.

TEAL version 3 (TEAL3), the most recent update to TEAL, is not Turing-complete as it does not support backwards branching. Although this restricts its functionality, it makes smart contracts written in TEAL more secure and easier to validate. Currently TEAL3 does not support functions either. There are only two types: uint64 and []byte in TEAL, TealType.uint64 and TealType.bytes in PyTeal.

Fig \ref{fig:pyteal} is an example of some code written in PyTeal and TEAL. The contract verifies that you are sending at least 50,000 microAlgos to a given address and the total spend including the transaction fee is less than 55,000 microAlgos. You can see the similarities between the two languages.

\begin{figure}[H]
\noindent\begin{minipage}{.55\textwidth}
\begin{lstlisting}[caption=PyTeal,frame=tlrb,language=Python]{Name}
from pyteal import *


def contract():
  to = Txn.receiver() == Addr(...)
  send = Txn.amount() > Int(50000)
  tot = Txn.amount() + Txn.fee()
  spend = tot < Int(55000)
  return And(
    to,
    send,
    spend
  )
\end{lstlisting}
\end{minipage}\hfill
\begin{minipage}{.41\textwidth}
\begin{lstlisting}[caption=TEAL,frame=tlrb]{Name}
txn Receiver
addr ...
==
txn Amount
int 50000
>
&&
txn Amount
txn Fee
+
int 55000
<
&&
\end{lstlisting}
\end{minipage}
\centering
\caption{Snippet of equivalent PyTeal and TEAL code}
\label{fig:pyteal}
\end{figure}

Algorand does not use the concept of gas to restrict TEAL program computation; there are no additional costs (on top of the fee that every transaction has) when calling an application irrespective of what gets executed. However there are some hard limits which restricts the size and contents of the TEAL programs you can write. 

TEAL programs: stateless and stateful are limited to 1,000 bytes and 1,024 bytes respectively. Each TEAL opcode also has an associated numeric value representing its cost. For many opcodes such as \textit{>}, the cost is 1, but for more computationally expensive opcodes, the cost is higher. To ensure a high transaction throughput there is a maximum opcode cost for TEAL programs. As stateless smart contracts live off-chain their total opcode cost limit is 20,000, whereas a stateful smart contract have a total opcode cost limit of 700.  


\subsection{Fees}
\label{subsection:fees}
All fees in Algorand are paid using Algos. The lowest denomination of Algos is microAlgos (1,000,000 microAlgos in an Algo). Every account in Algorand must have a minimum balance of 100,000 microAlgos. If a transaction is sent that would result in a balance below 100,000 microAlgos, then the transaction will fail. The minimum balance requirement applies to contract accounts as well.

There are fees for: submitting a transaction, opting into an ASA, creating a stateful smart contract and opting into a stateful smart contract. 

Transactions in Algorand have a 1,000 microAlgos fee regardless of the transaction type. If the number of transactions increase above the maximum supported 1,000 transactions per second (approximately) then these costs will increase as you compete with others to get your transaction recorded. However this has yet to happen for any sustained period of time.

The minimum balance requirement increases for every opted into ASA, and for creating and opting into a stateful smart contract. For an ASA the increase is 100,000 microAlgos. For stateful smart contracts the increase is determined by the following formula:

\[ 100000 + 285000 \times schema.NumUint + 50000 \times schema.NumByteSlice \]

For the creator, the variables refer to the global state and for the account that is opting in, the variables refer to the local state.

\chapter{Requirements and Design}
\label{chapter:requirementsanddesign}
In this chapter we will outline the green bond life cycle and the functionality needed from the stakeholders involved. We will then see how the solution meets these requirements and how the components interact.

\section{Stakeholders}
There are a number of stakeholders to consider when it comes to green bonds:
\begin{itemize}
    \item \textbf{Issuer} The issuer is the entity that borrows money by selling the green bond. This could be a government, bank or corporation.  
    \item \textbf{Investor} The investor is the entity that lends money by purchasing the green bond. This could be institutional or retail investors.
    \item \textbf{Green verifier} The agency that assesses the \textit{greenness} of the bond.
    \item \textbf{Financial regulator} The agency that keeps track that the bond is compliant with financial regulation.
\end{itemize}

\subsection{Issuer}
The issuer must be able to issue the green bond to the market and in return receive stablecoin payment. They would then be able to freely use these funds for their project. There must also be a mechanism for the issuer to fulfil the green bond requirements as per the green bond standards framework, and the ability to repay investors in the form of coupons and principal payments.

\subsection{Investor}
The investor must be able to purchase the available green bonds either directly from the issuer at the time of issuance or in the secondary market. They must be able to make informed decisions regarding the credit worthiness of the issuer, and the \textit{greenness} of the bond. Bond holders should also be able to receive regular coupon and principal payments using stablecoin.

\subsection{Green verifier}
The green verifier must be able to provide a green rating to a green bond initial use of proceeds proposal and also evaluate the ongoing green bond reporting of the issuer. To do this they will have to be able to read the issuers report and publicly share their assessments.

\subsection{Financial regulator}
The financial regulator must be able to approve the green bond issuance, monitor the financial transactions of the bond, freeze the bond in the presence of financial irregularities and also be able to prevent specific investors from benefiting from the bonds in cases such as AML violations.

\section{Bond Life Cycle}
\begin{figure}[H]
\includegraphics[scale=0.69]{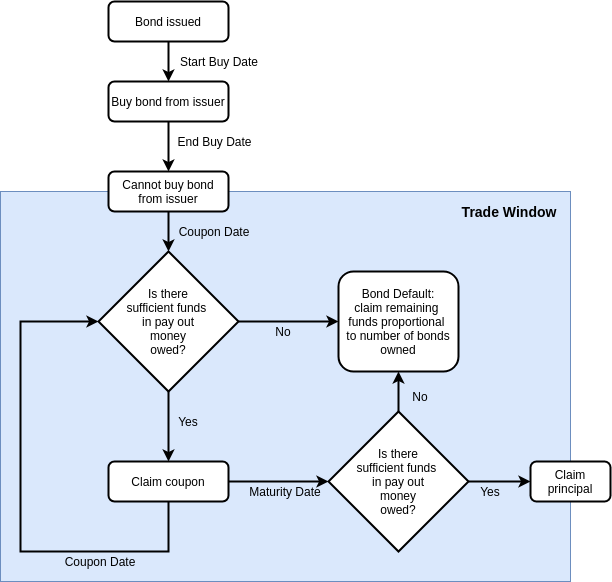}
\centering
\caption{The life cycle of a green bond}
\label{fig:lifecycle}
\end{figure}

The design for the life cycle of the green bond is very similar to that of a regular bond as shown in Fig \ref{fig:lifecycle}. 

The bond is initially issued and there is a set window where investors can buy the bond directly from the issuer at a fixed cost. 

After the buy period closes, the bond can be traded freely in the secondary market for the rest of its lifetime. There are periodic coupon payments specified according to the bond's configuration. At maturity the bond can be exchanged for its' principal.

At every payment event, there is a check to see whether the issuer has supplied enough funds to pay out all the money owed at that time. It is the responsibility of the issuer to transfer funds to an escrow account from which investors can claim their money from. For example if the coupon payments where \$10 per bond and there was 1,000 bonds in circulation, then the total money owed at the first coupon payment will be \$10,000. Similar checks are made for every coupon payment and the final principal payment. In the scenario where there is insufficient funds to pay the money owed then the bond defaults. At this point investors can claim a proportion of the remaining funds available that was from the issuer. For example, using the previous case where \$10,000 is owed and there is 1,000 bonds in circulation: if an investor owns 5 bonds and there is \$3,000 available then they would be able to claim \$15 of the \$50 owed, or 0.05\% of \$3,000.

\section{Green Rating}
\label{section:greenrating}
One of the key barriers to green bond adoption is a lack of green credit ratings \cite{greenbondbarriers}. Incorporating a rating mechanism into the structure of the green bond should help solve this issue.

\begin{figure}[H]
\includegraphics[scale=0.67]{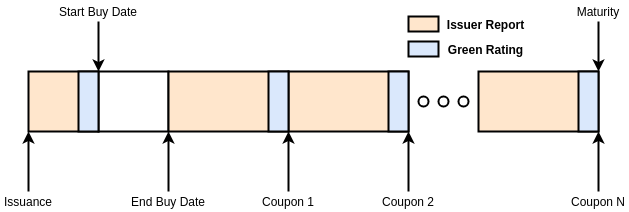}
\centering
\caption{The timeline of green reports and ratings}
\label{fig:reportratingtimeline}
\end{figure}

The green verifier provides ratings for the following events:
\begin{itemize}
    \item \textbf{Use of Proceeds} When the bond is first issued.
    \item \textbf{Reporting} At a coupon payment period, typically annually or biannually. 
\end{itemize}
The rating is on a scale between one and five, one being the worst and five being the best. Figure \ref{fig:reportratingtimeline} summarises the report-rating timeline.

Between issuance and the date from which investors can purchase the bond, the issuer submits a report for the \textit{use of proceeds} of the bond. The green verifier can use the uploaded document(s) to then assign a green rating which must also be done before the buy period.

In addition, at every coupon period there is an opportunity for investors to add further reporting based on their green Key Performance Indicators (KPIs). Again the green verifier provides a green rating towards the end of the period based on this information.

Environmental, Social, and Governance (ESG) funds and more generally green investors are looking to place their money in green bonds which meet certain green standards and avoid \textit{greenwashing}. Low green ratings may result in reduced bond pricing in the secondary market and would reflect badly on the issuer and the ESG fund managers that have chosen to invest in a poorly rated green bond.

To increase incentives for issuers to meet their green targets and achieve high green ratings, Dr Enrico Biffis, an Associate Professor of Actuarial Finance at Imperial College Business School, suggested linking the bond coupon rate to the green rating. Similar approaches have been suggested in the past for incorporating carbon pricing into green bonds \cite{carbonprice}.

The design is such that a most recent rating of 5 means that the coupon payment specified at issuance is used to determine the coupon payouts to investors. However the issuer is penalised for lower ratings, where each drop in rating increases the coupon payments by 10\%. Table \ref{tab:couponratinglink} provides an example of how this would work.

\begingroup
\renewcommand{\arraystretch}{1.5}
\begin{table}[H]
    \centering
    \begin{tabular}{c|c|c}
    \textbf{Green Rating} & \textbf{Penalty} & \textbf{Coupon Payment Value (2dp)} \\ \hline
    5                     & 0\%              & \$5.00                              \\
    4                     & 10\%             & \$5.50                              \\
    3                     & 21\%             & \$6.05                              \\
    2                     & 33.1\%           & \$6.66                              \\
    1                     & 46.4\%           & \$7.32                             
    \end{tabular}
    \caption{Effect of green ratings on coupon payments}
    \label{tab:couponratinglink}
\end{table}
\endgroup

\section{Secondary Market}
The bond can be freely traded among investors in the secondary market. Pricing is fully determined by market forces. Investors may use some of the following when making decisions regarding a specific green bond: the uploaded reports from the investor, the green ratings of the bond, the money in the bond's escrow account and the parameters of the bond.

\chapter{Implementation}
\label{chapter:implementation}

        
In this chapter we will cover the implementation details for the project. We will talk about the interesting problems encountered when developing the smart contracts and discuss alternative implementations with the reasons why they were not chosen. Lastly we will give an overview of the DApp architecture, coupled with the main features of the web application. 
        
\section{Algorand Blockchain}
All the smart contracts and assets were deployed on the Algorand TestNet. The TestNet allows developers to test their applications with the latest protocol version before before deploying to the MainNet. There are faucets available so that one can fund their TestNet accounts with Algos to use for free.

The implementation made use of a variety of Algorand features which are covered in section \ref{section:algorand}. In particular: atomic transfers, ASAs, stateful smart contracts and stateless smart contracts (contract accounts and delegated signatures).

\subsection{Stablecoin}
In Algorand, stablecoins such as Tether USDt and USDC are simply ASAs. Therefore since we do not have a way of obtaining these for testing purposes, we created a new ASA on the blockchain with a total supply of 1,000,000,000,000 and 6 decimals. 

By distributing this stablecoin ASA to accounts, we can replicate the scenario of investors transferring stablecoin to issuers and other investors to purchase bonds, and also the associated green bond coupon and principal stablecoin payments.

\subsection{Green Bond}
\label{subsection:greenbond}
For each green bond issuance, a new ASA is created with a unique identification number. Its supply is equal to the number of bonds the issuers would like to sell and has 6 decimals to support fractional bond ownership. For instance, an investor with 0.5 of this ASA can exchange their holdings for half the principal amount at maturity.

\subsection{Linking Stateful and Stateless Smart Contracts}
\label{subsection:linkingstatefulandstatelesssmartcontracts}
Before detailing the smart contracts involved, it is important to understand how stateful and stateless smart contracts can be linked. As explained in sections \ref{subsection:statelesssmartcontracts} and \ref{subsection:statefulsmartcontracts}, stateless smart contracts are used to approve transactions spent from an account and stateful smart contracts are used to store and manage state. What happens then when you need to use on-chain data (stateful) to approve transactions from an account (stateless)? The answer is that you link together the two.

\begin{figure}[H]
\includegraphics[scale=0.9]{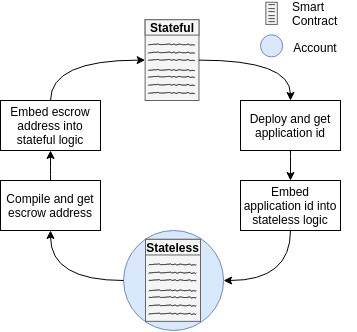}
\centering
\caption{Linking (stateless) contract account and stateful application}
\label{fig:linkingstatefulandstateless}
\end{figure}

When you deploy a stateful smart contract you get an an application identification number. The stateless smart contract can incorporate this into its logic by declaring that all transactions from the account will fail unless they are grouped with an application call to the stateful smart contract. The stateless smart contract can be compiled and you get the escrow address for the contract account. If you similarly want to require an application call to the stateful smart contract state to also submit a transaction from the stateless smart contract then you can embed its escrow address into its TEAL. However this creates a cycle where the stateful and stateless depend on one another as shown in figure \ref{fig:linkingstatefulandstateless}. To solve this issue you can add the escrow address to the already deployed application by either adding it to its state or using the \textit{UpdateApplication} transaction described in section \ref{subsection:statefulsmartcontracts}. We made use of the second approach.

\subsection{ASA Custom Transfer Logic}
\label{subsection:asacustomtransferlogic}
As discussed in section \ref{subsection:asa} assets can be traded similar to how Algos are traded, by submitting an asset transfer transaction. However by utilising the freeze and clawback functionality, one can require custom logic to be executed every time these assets are traded. This logic can track these assets as well as approve / reject any transactions which transfer them. 

\begin{figure}[H]
\includegraphics[scale=0.9]{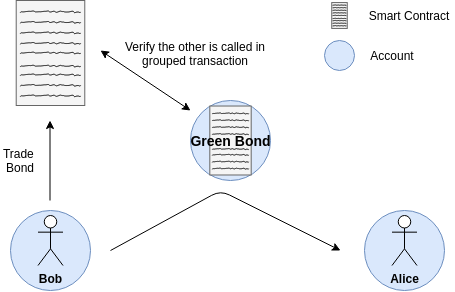}
\centering
\caption{Asset custom transfer using smart contract logic}
\label{fig:assetcustomtransferlogic}
\end{figure}

The figure \ref{fig:assetcustomtransferlogic} outlines how this would work for example when trading a bond in the secondary market. The ASA which represents the bond is frozen so they cannot be transferred freely between two parties. Frozen assets can still be transferred though by the clawback account which has the ability to revoke the asset holding of an account and send them to another account. We can therefore utilise stateless smart contract functionality, and set the clawback account to be a contract account. At this point the only way for the bond to be transferred is for the logic of the contract account to approve the transaction.

This solution provides the mechanism for the bond to be transferred from the issuer to the investor when they group the transaction atomically with a stablecoin payment and vice versa when the bond is exchanged for the principal at maturity. 

Furthermore we can also use on-chain data when approving these asset transfer transactions by linking the stateless smart contract with a stateful smart contract as explained in section \ref{subsection:linkingstatefulandstatelesssmartcontracts}. 

Fig \ref{fig:assetcustomtransferlogic} shows how all of this comes together. Let's say Bob wants to transfer the green bond in the secondary market to Alice. Bob cannot submit an \textit{AssetTransfer} transaction as the ASA representing the green bond is frozen. Bob must instead have the clawback account perform the transfer on their behalf. The clawback account is this case is a contract account; it is controlled by its TEAL logic. In this logic it outlines that Bob must call the stateful smart contract or otherwise it will reject the clawback transaction. At this point we finally have our solution: Bob calls the stateful smart contract, and groups that with a clawback transaction signed by the logic of the stateless smart contract.



\subsection{Smart Contracts}
For each green bond issuance, a new set of smart contracts are created and deployed:

\begin{itemize}
    \item \textbf{Main} Stateful smart contract that exposes mechanisms to buy green bonds, trade green bonds, claim coupons and claim principals.
    \item \textbf{Manage} Stateful smart contract that manages green ratings and bond defaults.
    \item \textbf{Bond Contract Account} Stateless smart contract whose logic approves bond transfers.
    \item \textbf{Stablecoin Contract Account} Stateless smart contract that issuer funds and whose logic approves stablecoin payments for coupons and principal.
\end{itemize}

In sections \ref{subsection:linkingstatefulandstatelesssmartcontracts} and \ref{subsection:asacustomtransferlogic} we outlined how stateless and stateful smart contracts can be linked together, and how to ensure all ASA transfers are approved / rejected by smart contract logic. In our case the \textit{Bond Contract Account} is the stateless contract account that is set as the clawback account for the green bond ASA. To approve any bond transfer (clawback transaction), it firsts checks that it is atomically grouped with an application call to the \textit{Main} application.  

The \textit{Main} application contains the following key value pairs in its storage:
\begin{itemize}
    \item \textbf{Local CouponsPaid} The number of collected coupons by an investor and their green bonds. 
    \item \textbf{Local Trade} The number of bonds owned by the account that are willing to be traded. This paired with a delegated signature, see section \ref{subsection:trade}, is used to facilitate trading in the secondary market.
    \item \textbf{Local Frozen} Boolean for if an account is frozen. Determines if they can buy / trade the green bonds and claim coupons / principal payments. Initially this is set to false (zero) such that only approved investors can purchase the green bond.
    \item \textbf{Global CouponsPaid} The maximum Local CouponsPaid across all investors. Used for bond defaults.
    \item \textbf{Global Reserve} The amount of stablecoin reserved in the stablecoin contract account that will be used to fund coupon and principal payments. Used for bond defaults.
    \item \textbf{Global Frozen} Boolean for if all accounts are frozen. Determines if anyone can buy / trade the green bonds and claim coupons / principal payments. Initially this is set to false (zero) such that only approved issuers can sell their green bonds.
\end{itemize}

The \textit{Manage} application contains the key value pairs for the green rating. Ideally we would like to store an array of length (No. Coupon Rounds + 1). Index 0 would contain the green rating for the use of proceeds and all the other postions will contain a green rating for the corresponding coupon round. However in PyTeal / TEAL there is no support for arrays. 

Instead we can use the key value map as an array. The key \textit{0} can contain the use of proceeds green rating, the key \textit{1} the first coupon green rating and so on. However this is costly as we would need one key value pair for every rating, and moreover the global state is limited to 64 key value pairs in total. Therefore because each value is 8 bytes in length, we store 8 ratings in each state value. Now we only need Math.ceil((No. Coupon Rounds + 1) / 8) key value pairs.

For example a green bond which has ratings of 5, 3, 4, 5, 5, 5, 5, 4, 3 would have state:
\begin{itemize}
    \item \textbf{Global 0} 0x0503040505050504 base16
    \item \textbf{Global 1} 0x0300000000000000 base16
\end{itemize}

One interesting implementation decision when writing the stateful smart contracts concerned when to hard code values in the TEAL smart contracts and when to rather store them in the global state. The advantage of hard coding the values is that there is no extra cost of having extra key value pairs in the global state. However this means that it is difficult to read these values from the blockchain as the smart contracts have to be decompiled and then parsed. In contrast, if the values are in the global state, they can be easily read from the blockchain but this incurs an added storage cost.

The approach chosen was that any values that are updated during the bond life cycle is stored in the smart contract state and all other values that are set at issuance and hard coded into the logic. For example the number of coupon rounds paid in stored in the smart contract state whereas the maturity date is not.

\textit{Main} is a stateful smart contract which coordinates everything together. A typical pattern one can use in an application in Algorand is to pass a string to a \textit{NoOp} application call and execute different logic depending on the string. This gives the impression of an application having different methods an account can call.

The \textit{Main} application supports passing the following strings:
\begin{itemize}
    \item \textbf{freeze}
    \item \textbf{freeze\_all}
    \item \textbf{buy}
    \item \textbf{set\_trade}
    \item \textbf{trade}
    \item \textbf{coupon}
    \item \textbf{sell}
    \item \textbf{default}
\end{itemize}

In addition the \textit{Manage} application supports passing the following strings:
\begin{itemize}
    \item \textbf{rate} 
    \item \textbf{defaulted}
    \item \textbf{not\_defaulted}
    \item \textbf{claim\_default}
\end{itemize}

We will cover these in more detail in the following sections.

\subsection{Green Rating}
Fig \ref{fig:reportratingtimeline} gives an overview of the timeline for green verifier ratings. The green verifier is able to call the \textit{Manage} application with the string "rate". The transaction also requires a second argument for the rating which is an integer between one and five inclusive. The time that the transaction was submitted at will determine what the rating is for.

\begingroup
\renewcommand{\arraystretch}{1.2}
\begin{table}[H]
    \centering
    \begin{adjustbox}{max width=\textwidth}
    \begin{tabular}{c|c|c|c|c|c|c|c}
    \textbf{Tx} & \textbf{Type}                                           & \textbf{Sign}                                        & \textbf{From}                                            & \textbf{App} & \textbf{Arg0} & \textbf{Arg1}                   \\ \hline
    \textbf{0}  & \begin{tabular}[c]{@{}c@{}}NoOp\\ App Call\end{tabular} & \begin{tabular}[c]{@{}c@{}}Secret\\ Key\end{tabular} & \begin{tabular}[c]{@{}c@{}}Green\\ Verifier\end{tabular} & Manage     & rate       & \textless{}rating\textgreater{}
    \end{tabular}
    \end{adjustbox}
    \caption{Rate transaction}
    \label{tab:ratetransaction}
\end{table}
\endgroup

\subsection{Freeze}
The financial regulator can call the \textit{Main} application with either the string "freeze" or "freeze\_all". 

The green bond is default frozen until the financial regulator approves the listing. They do this using "freeze\_all" and passing a non-zero number as an argument, and can also later revoke this license by calling "freeze\_all" with argument zero.

\begingroup
\renewcommand{\arraystretch}{1.2}
\begin{table}[H]
    \centering
    \begin{adjustbox}{max width=\textwidth}
    \begin{tabular}{c|c|c|c|c|c|c}
    \textbf{Tx} & \textbf{Type}                                           & \textbf{Sign}                                        & \textbf{From}                                                 & \textbf{App} & \textbf{Arg1} & \textbf{Arg2}                   \\ \hline
    \textbf{0}  & \begin{tabular}[c]{@{}c@{}}NoOp\\ App Call\end{tabular} & \begin{tabular}[c]{@{}c@{}}Secret\\ Key\end{tabular} & \begin{tabular}[c]{@{}c@{}}Financial\\ Regulator\end{tabular} & Main         & freeze\_all   & \textless{}number\textgreater{}
    \end{tabular}
    \end{adjustbox}
    \caption{Freeze All Transaction}
    \label{tab:freezeall}
\end{table}
\endgroup

The financial regulator can also freeze a specific account from interacting with the green bond, whether buying, trading or claiming money on their already owned bonds. Table \ref{tab:freeze} outlines the freeze account transaction. Accounts are set to frozen by default which allows the financial regulator to perform KYC and AML checks before approving an investor.

\begingroup
\renewcommand{\arraystretch}{1.2}
\begin{table}[H]
    \centering
    \begin{adjustbox}{max width=\textwidth}
    \begin{tabular}{c|c|c|c|c|c|c}
    \textbf{Tx} & \textbf{Type}                                           & \textbf{Sign}                                        & \textbf{From}                                                 & \textbf{App} & \textbf{Arg1} & \textbf{Arg2}                   \\ \hline
    \textbf{0}  & \begin{tabular}[c]{@{}c@{}}NoOp\\ App Call\end{tabular} & \begin{tabular}[c]{@{}c@{}}Secret\\ Key\end{tabular} & \begin{tabular}[c]{@{}c@{}}Financial\\ Regulator\end{tabular} & Main         & freeze        & \textless{}number\textgreater{}
    \end{tabular}
    \end{adjustbox}
    \caption{Freeze Account Transaction}
    \label{tab:freeze}
\end{table}
\endgroup

\subsection{Buy}
You can buy the green bond between the start buy date and end buy date by submitting the transactions in Table \ref{tab:buytransactions} grouped together. The bond and stablecoin payment are exchanged using the asset custom transfer logic we spoke about earlier in section \ref{subsection:asacustomtransferlogic}.

\begingroup
\renewcommand{\arraystretch}{1.2}
\begin{table}[H]
    \centering
    \begin{adjustbox}{max width=\textwidth}
    \begin{tabular}{c|c|c|c|c|c|c|c|c}
    \textbf{Tx} & \textbf{Type}                                                           & \textbf{Sign}                                             & \textbf{From}                                         & \textbf{\begin{tabular}[c]{@{}c@{}}Revoke\\ Target\end{tabular}} & \textbf{To}                                           & \textbf{Amount}                                      & \textbf{App} & \textbf{Args} \\ \hline
    \textbf{0}  & \begin{tabular}[c]{@{}c@{}}NoOp\\ App Call\end{tabular}                 & \begin{tabular}[c]{@{}c@{}}Secret\\ Key\end{tabular}      & Investor                                              & -                                                                    & -                                                     & -                                                    & Main         & buy           \\ \hline
    \textbf{1}  & \begin{tabular}[c]{@{}c@{}}Algo\\ Transfer\end{tabular}                 & \begin{tabular}[c]{@{}c@{}}Secret\\ Key\end{tabular}      & Investor                                              & -                                                                    & \begin{tabular}[c]{@{}c@{}}Bond\\ Escrow\end{tabular} & \begin{tabular}[c]{@{}c@{}}Fee\\ Of Tx2\end{tabular} & -            & -             \\ \hline
    \textbf{2}  & \begin{tabular}[c]{@{}c@{}}Asset\\ Transfer\\ (Bond)\end{tabular}       & \begin{tabular}[c]{@{}c@{}}Logic\\ Signature\end{tabular} & \begin{tabular}[c]{@{}c@{}}Bond\\ Escrow\end{tabular} & \begin{tabular}[c]{@{}c@{}}Bond\\ Escrow\end{tabular}                & Investor                                              & N                                                    & -            & -             \\ \hline
    \textbf{3}  & \begin{tabular}[c]{@{}c@{}}Asset\\ Transfer\\ (Stablecoin)\end{tabular} & \begin{tabular}[c]{@{}c@{}}Secret\\ Key\end{tabular}      & Investor                                              & -                                                                    & Issuer                                                & N*Cost                                               & -            & -            
    \end{tabular}
    \end{adjustbox}
    \caption{Buy Atomic Transaction Group}
    \label{tab:buytransactions}
\end{table}
\endgroup

\subsection{Trade}
\label{subsection:trade}
To trade the green bond, the owner must call the \textit{Main} application with "trade". This must be coupled with the bond transfer, using the clawback functionality, and also a small fee for bond escrow that revoked the asset from investor 1 to investor 2. 

The trade verification makes no assumption about any further transactions in the group. For example one may decide to gift their green bond to a friend or sell it on the secondary market for Algos. 

Imagine an investor wants to sell their green bond on the secondary market in exchange for stablecoin. The blockchain can facilitate this transfer without the need for an intermediary, ensuring both the bond holder and buyer can trust the transfer.

The bond holder first sets the number of bonds they are willing to trade using the transaction in table \ref{tab:settradetransaction}. Then a stateless smart contract can be generated which specifies the terms under which they would sell the bond - the stablecoin amount and an expiry date. This is then signed with the investor's private key and can now be used as a delegated signature by the buyer. The buyer can submit the following group of transactions in table \ref{tab:tradetransactions} to trigger the trade of the bond.

\begingroup
\renewcommand{\arraystretch}{1.2}
\begin{table}[H]
    \centering
    \begin{adjustbox}{max width=\textwidth}
    \begin{tabular}{c|c|c|c|c|c|c}
    \textbf{Tx} & \textbf{Type}                                           & \textbf{Sign}                                        & \textbf{From}                                        & \textbf{App} & \textbf{Arg0}                                       & \textbf{Arg1} \\ \hline
    \textbf{0}  & \begin{tabular}[c]{@{}c@{}}NoOp\\ App Call\end{tabular} & \begin{tabular}[c]{@{}c@{}}Secret\\ Key\end{tabular} & \begin{tabular}[c]{@{}c@{}}Investor\\ 1\end{tabular} & Main         & \begin{tabular}[c]{@{}c@{}}set\\ trade\end{tabular} & N           
    \end{tabular}
    \end{adjustbox}
    \caption{Set number of bonds willing to trade}
    \label{tab:settradetransaction}
\end{table}
\endgroup

The reason why the bond holder first sets in their local state how many bonds they would like to trade in total is because otherwise people can take advantage of their delegated signature. For example if an investor were to encode in a stateless smart contract that they would like to sell 2 bonds then that delegated signature could be used multiple times until the investor no longer owns any green bonds. The stateless smart contract logic has no concept of on-chain data so cannot tell whether it has already been used before when approving future trades. Subsequently we use the bond holders local storage and if they are willing to sell 2 of their bonds at a cost of \$1,000 per bond then a person can come along and buy 0.5 bonds for \$500 which would update the state value to 1.5 bonds. Then a second buyer could purchase the remaining 1.5 bonds at \$1,500 and then at that point all trades from that account will be blocked by the \textit{Main} application call (or transaction 0).

\begingroup
\renewcommand{\arraystretch}{1.2}
\begin{table}[H]
    \centering
    \begin{adjustbox}{max width=\textwidth}
    \begin{tabular}{c|c|c|c|c|c|c|c|c}
    \textbf{Tx} & \textbf{Type}                                                           & \textbf{Sign}                                             & \textbf{From}                                         & \textbf{\begin{tabular}[c]{@{}c@{}}Revoke\\ Target\end{tabular}} & \textbf{To}                                           & \textbf{Amount}                                      & \textbf{App} & \textbf{Args} \\ \hline
    \textbf{0}  & \begin{tabular}[c]{@{}c@{}}NoOp\\ App Call\end{tabular}                 & \begin{tabular}[c]{@{}c@{}}Logic\\ Signature\end{tabular} & \begin{tabular}[c]{@{}c@{}}Investor\\ 1\end{tabular}  & -                                                                & -                                                     & -                                                    & Main         & trade         \\ \hline
    \textbf{1}  & \begin{tabular}[c]{@{}c@{}}Algo\\ Transfer\end{tabular}                 & \begin{tabular}[c]{@{}c@{}}Logic\\ Signature\end{tabular} & \begin{tabular}[c]{@{}c@{}}Investor\\ 1\end{tabular}  & -                                                                & \begin{tabular}[c]{@{}c@{}}Bond\\ Escrow\end{tabular} & \begin{tabular}[c]{@{}c@{}}Fee\\ Of Tx2\end{tabular} & -            & -             \\ \hline
    \textbf{2}  & \begin{tabular}[c]{@{}c@{}}Asset\\ Transfer\\ (Bond)\end{tabular}       & \begin{tabular}[c]{@{}c@{}}Logic\\ Signature\end{tabular} & \begin{tabular}[c]{@{}c@{}}Bond\\ Escrow\end{tabular} & \begin{tabular}[c]{@{}c@{}}Investor\\ 1\end{tabular}             & \begin{tabular}[c]{@{}c@{}}Investor\\ 2\end{tabular}  & N                                                    & -            & -             \\ \hline
    \textbf{3}  & \begin{tabular}[c]{@{}c@{}}Asset\\ Transfer\\ (Stablecoin)\end{tabular} & \begin{tabular}[c]{@{}c@{}}Secret\\ Key\end{tabular}      & \begin{tabular}[c]{@{}c@{}}Investor\\ 2\end{tabular}  & -                                                                & \begin{tabular}[c]{@{}c@{}}Investor\\ 1\end{tabular}  & N*Price                                              & -            &              
    \end{tabular}
    \end{adjustbox}
    \caption{Trade Atomic Transaction Group Using Logic Signature}
    \label{tab:tradetransactions}
\end{table}
\endgroup

\subsection{Coupon}
Green bond owners can claim coupons using the transactions in table \ref{tab:coupontransactions}. As explained in the requirements section \ref{section:greenrating}, the amount of stablecoin paid is inversely proportional to the green rating of the bond. The smart contract calls verify that the correct payment is made based on this rating.

\begingroup
\renewcommand{\arraystretch}{1.2}
\begin{table}[H]
    \centering
    \begin{adjustbox}{max width=\textwidth}
    \begin{tabular}{c|c|c|c|c|c|c|c}
    \textbf{Tx} & \textbf{Type}                                                           & \textbf{Sign}                                             & \textbf{From}                                               & \textbf{To}                                                 & \textbf{Amount}                                      & \textbf{App} & \textbf{Args}                                           \\ \hline
    \textbf{0}  & \begin{tabular}[c]{@{}c@{}}NoOp\\ App Call\end{tabular}                 & \begin{tabular}[c]{@{}c@{}}Secret\\ Key\end{tabular}      & Investor                                                    & -                                                           & -                                                    & Main         & coupon                                                  \\ \hline
    \textbf{1}  & \begin{tabular}[c]{@{}c@{}}NoOp\\ App Call\end{tabular}                 & \begin{tabular}[c]{@{}c@{}}Secret\\ Key\end{tabular}      & Investor                                                    & -                                                           & -                                                    & Manage       & \begin{tabular}[c]{@{}c@{}}not\\ defaulted\end{tabular} \\ \hline
    \textbf{2}  & \begin{tabular}[c]{@{}c@{}}Algo\\ Transfer\end{tabular}                 & \begin{tabular}[c]{@{}c@{}}Secret\\ Key\end{tabular}      & Investor                                                    & \begin{tabular}[c]{@{}c@{}}Stablecoin\\ Escrow\end{tabular} & \begin{tabular}[c]{@{}c@{}}Fee\\ Of Tx3\end{tabular} & -            & -                                                       \\ \hline
    \textbf{3}  & \begin{tabular}[c]{@{}c@{}}Asset\\ Transfer\\ (Stablecoin)\end{tabular} & \begin{tabular}[c]{@{}c@{}}Logic\\ Signature\end{tabular} & \begin{tabular}[c]{@{}c@{}}Stablecoin\\ Escrow\end{tabular} & Investor                                                    & N*Coupon                                             & -            & -                                                      
    \end{tabular}
    \end{adjustbox}
    \caption{Claim Coupon Atomic Transaction Group}
    \label{tab:coupontransactions}
\end{table}
\endgroup

One interesting consideration is checking if the green bond has defaulted. Each time someone tries to claim a coupon, we use the local number of coupon payments collected and compare that to the global maximum number of coupon payments across all investors. If the transactions are the first time a coupon has been claimed for that round, which we know when \textit{Local CouponsPaid} exceeds \textit{Global CouponsPaid}, then we need to check that the bond has not defaulted. This is done by incrementing \textit{Global Reserve} by the coupon payment value multiplied by the number of bonds in circulation. We then assert in \textit{Manage} application that there is enough funds in the stablecoin escrow amount to meet the entire \textit{Reserve} amount. 

Table \ref{tab:coupondefaultrunthrough} is an example run through of how this works in practise. We are using a fixed \$100 coupon payment per bond to simplify the calculations here but in reality this will be dependant on the green ratings.

\begingroup
\renewcommand{\arraystretch}{1.2}
\begin{table}[H]
    \centering
    \begin{adjustbox}{max width=\textwidth}
    \begin{tabular}{c|c|c|c|c|c|c|c}
    \textbf{Action}                                                          & \textbf{\begin{tabular}[c]{@{}c@{}}No. Bonds\\ In \\ Circulation\end{tabular}} & \textbf{\begin{tabular}[c]{@{}c@{}}Invetor 1\\ Local\\ Coupons\\ Payed\end{tabular}} & \textbf{\begin{tabular}[c]{@{}c@{}}Invetor 2\\ Local\\ Coupons\\ Payed\end{tabular}} & \textbf{\begin{tabular}[c]{@{}c@{}}Global\\ Coupons\\ Payed\end{tabular}} & \textbf{\begin{tabular}[c]{@{}c@{}}Reserve\\ (\$)\end{tabular}} & \textbf{\begin{tabular}[c]{@{}c@{}}Stablecoin \\ Escrow \\ Balance (\$)\end{tabular}} & \textbf{Approved?}       \\ \hline
    \begin{tabular}[c]{@{}c@{}}Investor 1 buys\\ 5 green bonds\end{tabular}  & 5                                                                              & 0                                                                                    & 0                                                                                    & 0                                                                         & {\color[HTML]{009901} 0}                                        & 0                                                                                     & \multicolumn{1}{c|}{yes} \\ \hline
    \begin{tabular}[c]{@{}c@{}}Investor 2 buys\\ 10 green bonds\end{tabular} & 15                                                                             & 0                                                                                    & 0                                                                                    & 0                                                                         & {\color[HTML]{009901} 0}                                        & 0                                                                                     & \multicolumn{1}{c|}{yes} \\ \hline
    \begin{tabular}[c]{@{}c@{}}Issuer funds \\ escrow \$1,500\end{tabular}   & 15                                                                             & 0                                                                                    & 0                                                                                    & 0                                                                         & {\color[HTML]{009901} 0}                                        & 1,500                                                                                 & \multicolumn{1}{c|}{yes} \\ \hline
    \begin{tabular}[c]{@{}c@{}}Investor 2\\ claims coupon\end{tabular}       & 15                                                                             & 0                                                                                    & 1                                                                                    & 1                                                                         & {\color[HTML]{009901} 500}                                      & 500                                                                                   & \multicolumn{1}{c|}{yes} \\ \hline
    \begin{tabular}[c]{@{}c@{}}Investor 1\\ claims coupon\end{tabular}       & 15                                                                             & 1                                                                                    & 1                                                                                    & 1                                                                         & {\color[HTML]{009901} 0}                                        & 0                                                                                     & \multicolumn{1}{c|}{yes} \\ \hline
    \begin{tabular}[c]{@{}c@{}}Issuer funds \\ escrow \$1,000\end{tabular}   & 15                                                                             & 1                                                                                    & 1                                                                                    & 1                                                                         & {\color[HTML]{009901} 0}                                        & 1000                                                                                  & \multicolumn{1}{c|}{yes} \\ \hline
    \begin{tabular}[c]{@{}c@{}}Investor 1\\ claims coupon\end{tabular}       & 15                                                                             & 2                                                                                    & 1                                                                                    & 2                                                                         & {\color[HTML]{FE0000} 1500}                                     & 1000                                                                                  & \multicolumn{1}{c|}{no} 
    \end{tabular}
    \end{adjustbox}
    \caption{Default checks when investors claiming \$100 coupons}
    \label{tab:coupondefaultrunthrough}
\end{table}
\endgroup

\subsection{Principal}
Claiming the green bond principal is similar to claiming coupons with the addition of forfeiting all bonds owned. At maturity, the investor can submit the transactions in table \ref{tab:principaltransactions} atomically grouped.

In addition to verifying there is sufficient funds to pay all coupons owed, there is a check to ensure all principals can be payed. If there is not enough money to do so, then the grouped transactions will be rejected.

\begingroup
\renewcommand{\arraystretch}{1.2}
\begin{table}[H]
    \centering
    \begin{adjustbox}{max width=\textwidth}
    \begin{tabular}{c|c|c|c|c|c|c|c|c}
    \textbf{Tx} & \textbf{Type}                                                           & \textbf{Sign}                                             & \textbf{From}                                               & \textbf{\begin{tabular}[c]{@{}c@{}}Revoke\\ Target\end{tabular}} & \textbf{To}                                                 & \textbf{Amount}                                      & \textbf{App} & \textbf{Args}                                           \\ \hline
    \textbf{0}  & \begin{tabular}[c]{@{}c@{}}NoOp\\ App Call\end{tabular}                 & \begin{tabular}[c]{@{}c@{}}Secret\\ Key\end{tabular}      & Investor                                                    & -                                                                & -                                                           & -                                                    & Main         & coupon                                                  \\ \hline
    \textbf{1}  & \begin{tabular}[c]{@{}c@{}}NoOp\\ App Call\end{tabular}                 & \begin{tabular}[c]{@{}c@{}}Secret\\ Key\end{tabular}      & Investor                                                    & -                                                                & -                                                           & -                                                    & Manage       & \begin{tabular}[c]{@{}c@{}}not\\ defaulted\end{tabular} \\ \hline
    \textbf{2}  & \begin{tabular}[c]{@{}c@{}}Asset\\ Transfer\\ (Bond)\end{tabular}       & \begin{tabular}[c]{@{}c@{}}Logic\\ Signature\end{tabular} & \begin{tabular}[c]{@{}c@{}}Bond\\ Escrow\end{tabular}       & Investor                                                         & \begin{tabular}[c]{@{}c@{}}Bond\\ Escrow\end{tabular}       & N                                                    & -            & -                                                       \\ \hline
    \textbf{3}  & \begin{tabular}[c]{@{}c@{}}Asset\\ Transfer\\ (Stablecoin)\end{tabular} & \begin{tabular}[c]{@{}c@{}}Logic\\ Signature\end{tabular} & \begin{tabular}[c]{@{}c@{}}Stablecoin\\ Escrow\end{tabular} & -                                                                & Investor                                                    & N*Principal                                          & -            & -                                                       \\ \hline
    \textbf{4}  & \begin{tabular}[c]{@{}c@{}}Algo\\ Transfer\end{tabular}                 & \begin{tabular}[c]{@{}c@{}}Secret\\ Key\end{tabular}      & Investor                                                    & -                                                                & \begin{tabular}[c]{@{}c@{}}Bond\\ Escrow\end{tabular}       & \begin{tabular}[c]{@{}c@{}}Fee\\ Of Tx2\end{tabular} & -            & -                                                       \\ \hline
    \textbf{5}  & \begin{tabular}[c]{@{}c@{}}Algo\\ Transfer\end{tabular}                 & \begin{tabular}[c]{@{}c@{}}Secret\\ Key\end{tabular}      & Investor                                                    & -                                                                & \begin{tabular}[c]{@{}c@{}}Stablecoin\\ Escrow\end{tabular} & \begin{tabular}[c]{@{}c@{}}Fee\\ Of Tx3\end{tabular} & -            & -                                                      
    \end{tabular}
    \end{adjustbox}
    \caption{Claim Principal Atomic Transaction Group}
    \label{tab:principaltransactions}
\end{table}
\endgroup

\subsection{Default}
To claim a default, there must be insufficient funds to pay all money owed to investors. The implementation first verifies that the bond owner has claimed all the coupon payments that they are entitled to up to and including the \textit{Global CouponsPayed}, ensuring investors do not lose out on any money based on their delayed stablecoin payment requests.

The \textit{Manage} application checks not whether a bond has been defaulted, but whether the one additional coupon payment (or principal) would cause the bond to default. That way investors are able to maximise the money they can recoup.

\begingroup
\renewcommand{\arraystretch}{1.2}
\begin{table}[H]
    \centering
    \begin{adjustbox}{max width=\textwidth}
    \begin{tabular}{c|c|c|c|c|c|c|c|c}
    \textbf{Tx} & \textbf{Type}                                                           & \textbf{Sign}                                             & \textbf{From}                                               & \textbf{\begin{tabular}[c]{@{}c@{}}Revoke\\ Target\end{tabular}} & \textbf{To}                                                 & \textbf{Amount}                                              & \textbf{App} & \textbf{Args}                                           \\ \hline
    \textbf{0}  & \begin{tabular}[c]{@{}c@{}}NoOp\\ App Call\end{tabular}                 & \begin{tabular}[c]{@{}c@{}}Secret\\ Key\end{tabular}      & Investor                                                    & -                                                                & -                                                           & -                                                            & Main         & default                                                 \\ \hline
    \textbf{1}  & \begin{tabular}[c]{@{}c@{}}NoOp\\ App Call\end{tabular}                 & \begin{tabular}[c]{@{}c@{}}Secret\\ Key\end{tabular}      & Investor                                                    & -                                                                & -                                                           & -                                                            & Manage       & \begin{tabular}[c]{@{}c@{}}claim\\ default\end{tabular} \\ \hline
    \textbf{2}  & \begin{tabular}[c]{@{}c@{}}Asset\\ Transfer\\ (Bond)\end{tabular}       & \begin{tabular}[c]{@{}c@{}}Logic\\ Signature\end{tabular} & \begin{tabular}[c]{@{}c@{}}Bond\\ Escrow\end{tabular}       & Investor                                                         & \begin{tabular}[c]{@{}c@{}}Bond\\ Escrow\end{tabular}       & N                                                            & -            & -                                                       \\ \hline
    \textbf{3}  & \begin{tabular}[c]{@{}c@{}}Asset\\ Transfer\\ (Stablecoin)\end{tabular} & \begin{tabular}[c]{@{}c@{}}Logic\\ Signature\end{tabular} & \begin{tabular}[c]{@{}c@{}}Stablecoin\\ Escrow\end{tabular} & -                                                                & Investor                                                    & \begin{tabular}[c]{@{}c@{}}Default\\ Proportion\end{tabular} & -            & -                                                       \\ \hline
    \textbf{4}  & \begin{tabular}[c]{@{}c@{}}Algo\\ Transfer\end{tabular}                 & \begin{tabular}[c]{@{}c@{}}Secret\\ Key\end{tabular}      & Investor                                                    & -                                                                & \begin{tabular}[c]{@{}c@{}}Bond\\ Escrow\end{tabular}       & \begin{tabular}[c]{@{}c@{}}Fee\\ Of Tx2\end{tabular}         & -            & -                                                       \\ \hline
    \textbf{5}  & \begin{tabular}[c]{@{}c@{}}Algo\\ Transfer\end{tabular}                 & \begin{tabular}[c]{@{}c@{}}Secret\\ Key\end{tabular}      & Investor                                                    & -                                                                & \begin{tabular}[c]{@{}c@{}}Stablecoin\\ Escrow\end{tabular} & \begin{tabular}[c]{@{}c@{}}Fee\\ Of Tx3\end{tabular}         & -            & -                                                      
    \end{tabular}
    \end{adjustbox}
    \caption{Claim Default Atomic Transaction Group}
    \label{tab:defaulttransactions}
\end{table}
\endgroup

When a bond defaults there are a number of options available to the investor. They can wait in the hope that the issuer transfers more stablecoin into the escrow account, can trade the bond in the secondary market to \textit{distressed debt} investors or exchange the bond for a proportion of the remaining funds in the escrow. For the last option, they submit the transaction group in table \ref{tab:defaulttransactions} and the stateful smart contracts together verify the stablecoin transfer is equal to:
\[ (StablecoinEscrowBalance - Global.Reserve) \times \frac{NumBondsOwned}{NumBondsInCirculation} \]

\subsection{InterPlanetary File System}
\label{subsection:ipfs}
As part of the green verifier rating, the issuer uploads documents for the use of proceeds and green reporting. Blockchain is not designed to store large amounts of data and it would be very costly to do so.

To overcome these limitations, we use the InterPlanetary File System (IPFS) to share the files uploaded by the issuer. IPFS is a protocol and peer-to-peer network for storing data in a distributed way. Each file is hashed to get a \textit{Content IDentifier} (CID) which can be used to uniquely refer to the file. When a client want to look up a particular file, all they need to do is supply the CID and the distributed nodes cooperate to return the contents from the global namespace.

To link this to the Algorand blockchain the issuer first uploads the file to IPFS and then submit a transaction from their Algorand account with the hash in the \textit{note} field. The issuer may be responsible for many bonds so the hash is prepended by the associated \textit{Manage} application identification number for the green bond. The structure is: 
\[ <MANAGE\_APP\_ID>+<IPFS\_CID> \]

When we want to retrieve the files for a given green bond, the blockchain can be searched by looking up all transactions from the issuers address that has the \textit{note} prefix of \textit{"<MANAGE\_APP\_ID>+"}. The suffix of the note for the transactions returned are the CIDs which can be used to query the distributed file system known as IPFS.

\subsection{Testing}
As Algorand smart contract development is relatively recent, there are not many tools available when it comes to testing. One method of testing smart contracts is by creating a local private network, deploying them there and manually checking their functionality. This is tedious and cumbersome as due to the time nature of the green bond life cycle, you have to wait until a certain coupon date or maturity. Furthermore all the setup takes several minutes.

The main way the implementation was tested was using the \textit{Algo Builder} framework \cite{algobuilder}. It includes a \textit{runtime} package which can simulate (most) Algorand transactions including stateful smart contract calls. This was combined with \textit{Mocha} for the test framework and \textit{Chai} for assertions. Together these were able to automate testing of all different scenarios and edge cases that would otherwise be difficult to test manually. 

\subsection{Alternative Implementations}
A number of different implementations were considered for the green bond. Section \ref{subsection:greenbond} outlined the decision to issue one ASA per green bond issuance with its supply equal to the number of bonds the issuer intends to sell. Two alternatives considered were:

\begin{enumerate}
    \item \textbf{ASA for every bond} Issuer wants to sell n number of bonds. In total n number of ASA are created, each with a supply of 1.
    \item \textbf{ASA for each coupon round and the final principal} Issuer wants to sell n number of bonds with p coupon payment periods (e.g. p=10 for 5 year bond with biannual coupons). In total p+1 ASAs are created, each with a supply of n.
\end{enumerate}

The benefits of the first approach is that you can distinguish between each bond using its unique identification number. Instead of storing in the application how many coupon payments were made to a particular account, you rather store how many coupon payments were made for a particular bond. You no longer need to track bond transfers in the secondary market as you will always be able to identify which bond is which.

There are a few problems with the first approach however, the most severe of which is that it does not scale. There is a fixed cost per ASA creation, you must opt in to receive an ASA which incurs a transaction fee, and currently a single Algorand account can opt into a maximum of 1,000 different ASA. This would restrict an account from owning more than 1,000 bonds. You can get around this by each user having more than one Algorand account when necessary but the costs are still way too high.

The second approach is a lot more viable than the first. Idea 2 enables investors to trade coupons independently of the bond. This is known as \textit{Coupon Bonds} or \textit{Bearer Bonds}. The stateful smart contract simply exchanges these coupon ASA for stablecoin, and coupon payments are no longer stored. 

One concern with this implementation is how the market will react to the new financial instrument. Pricing becomes different as you are not trading bonds but now a form of futures, which may deter some investors.

Another consideration one has to make is the cost of the solution. An investor who purchases the bond upfront would need to opt into every ASA. This is a significant upfront cost. In addition, there are opcode and size limitations for stateful smart contracts. One would need to store the asset identification numbers of many ASA which may breach these limits as the number of coupon payments grows. You could have one contract per ASA but this would be costly on the issuer.

\section{DApp}
A DApp was created as a demo application for how users would interact with the platform as a whole. The DApp gives the users the ability to act as an issuer, investor, green verifier or financial regulator. In practise, a user would have to be approved before they can have these permissions. For example an investor would have some KYC checks and only a handful of accounts would be a green verifier or financial regulator.

\begin{figure}[H]
\includegraphics[scale=0.57]{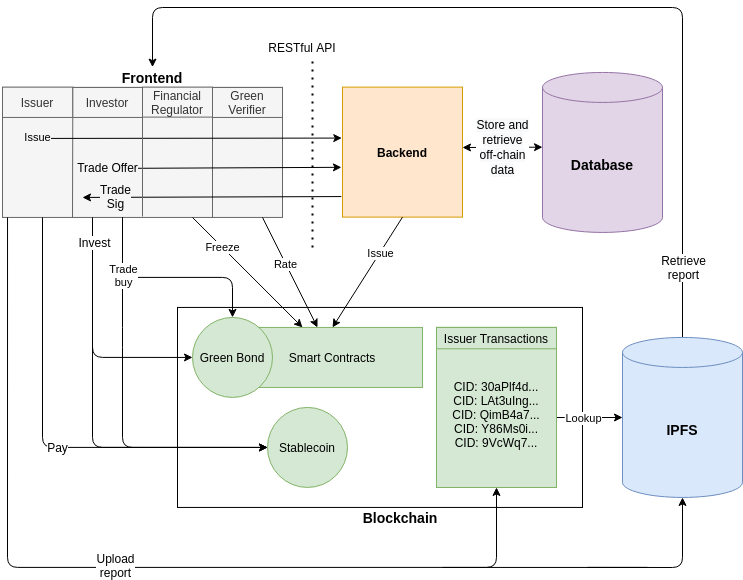}
\centering
\caption{DApp Architecture}
\label{fig:architecture}
\end{figure}

The DApp interacts with the Algorand blockchain using both a frontend user interface and backend server. The frontend allows users to sign transactions using their locally stored keys. The backend is used to interact with the blockchain using private keys which is used in issuance and the stablecoin dispenser. There is also a connected database for off-chain storage such as the addresses linked to a user account. The DApp architecture is shown in Figure \ref{fig:architecture}.

A user guide on how interact with the web application is provided in Appendix \ref{userguide}.


\subsection{Languages, Frameworks and Tools Used}
Some of the technologies used include: 

\begin{itemize}
    \item \textbf{TypeScript} \cite{typescript} Programming language that adds optional static typing to JavaScript. Used to write more reliable and easier to refactor code.
    \item \textbf{React} \cite{react} A JavaScript library for building user interfaces. Used to build encapsulated components, each with their own state, which can be combined to create complex views.
    \item \textbf{React Redux} \cite{reactredux} React bindings for Redux, a state management library for JavaScript applications. Used to share state across components.
    \item \textbf{Material-UI} \cite{material-ui} A React UI framework that provides components for faster and easier development. 
    \item \textbf{Node.js} \cite{nodejs} An asynchronous event-driven JavaScript runtime. Used in the server to handle concurrent network connections and requests.
    \item \textbf{Express} \cite{express} Node.js web framework. Used for routing and authentication middleware.
    \item \textbf{PostgreSQL} \cite{postgresql} Relational database system that uses the SQL language. Used to store off-chain data.
    \item \textbf{Auth0} \cite{auth0} Authentication platform. Used for authorisation and account management.
    \item \textbf{Rand Labs Algorand Developer API} \cite{algoranddeveloperapi} Used to interact with the blockchain without needing to run our own node.
    \item \textbf{MyAlgo Connect} \cite{myalgo-connect} JavaScript library to securely sign Algorand transaction with the MyAlgo wallet 
\end{itemize}

\subsection{Redux}
The Redux store holds the whole state tree of your application. To update the state inside it, you must dispatch an action. This triggers a call to the reducer which changes the state depending on the action and the previous state value.

For the application, the state tree is split into two branches: one to store state about the logged in user and the other to store state about the green bonds.

\begin{figure}[H]
\centering
\begin{lstlisting}
interface BondState {
  apps: Map<number, App>;
  selectedApp?: number;
  trades: Map<number, Trade>;
  selectedTrade?: number;
}

interface UserState {
  addresses: string[];
  selectedAccount?: UserAccount;
}
\end{lstlisting}
\caption{React Redux Store}
\label{fig:reduxstate}
\end{figure}

The \textit{BondState} interface contains two map data structures. The first links the main smart contract application identification number for a given green bond, to the application itself. The value contains information including the bond identification number, and blockchain readings like the stateful smart contracts' global state. The other map stores trade offers for a given application and will be discussed in more detail in section \ref{subsection:issuer}.

The \textit{UserState} interface is used to store the Algorand addresses connected to the user account and also information regarding the selected Algorand address. This comprises of blockchain readings for the Algo balance, owned green bonds and stateful smart contracts' local state.

\subsection{React Components}
The React components were divided into two categories: presentational components and container components. Presentational components control how the user interface looks and are often reusable, whereas container components manage state and the application logic. This design pattern helped separate to separate concerns, making it easier to refactor and reason about the code.

An example of the pattern in use are the components \textit{GreenVerifierPageContainer} and \textit{GreenVerifierPage}. As the name suggest, \textit{GreenVerifierPageContainer} is the container component. It utilises the \textit{connect()} function from Redux that allows the component to dispatch actions and read the global state. The component also contains the logic for submitting a green rating. \textit{GreenVerifierPageContainer} passes props to its child component \textit{GreenVerifierPage} which is the presentational component. \textit{GreenVerifierPage} is responsible for positioning the page components and does not contain any state.

\begin{figure}[H]
\includegraphics[scale=0.73]{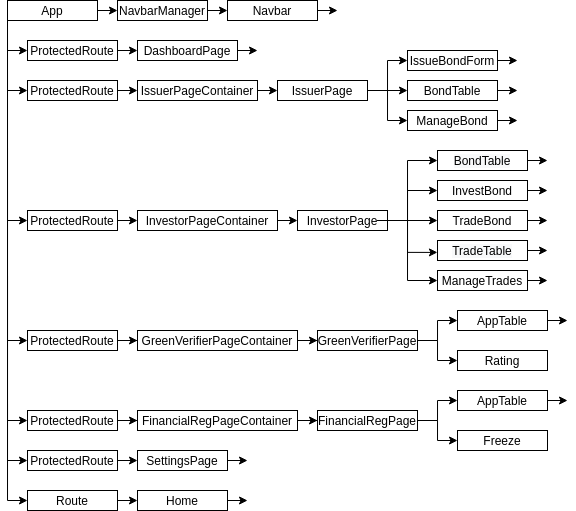}
\centering
\caption{Simplified React Component Tree}
\label{fig:reactcomponents}
\end{figure}

Figure \ref{fig:reactcomponents} shows a simplified component tree, starting from \textit{App}, the top level component. The \textit{NavbarManager} is shared across all pages so it is a direct child of \textit{App}. The React Router is used to simulate multiple URLs in the web application. It is coupled with Auth0 React package which protects each page so that only authenticated users can access them. If a user tries to go to one of these paths they are automatically redirected to the login page. Subsequently there are an additional six \textit{ProtectedRoute} children components for the dashboard, issuer, investor, green verifier, financial regulator and settings pages. The last component is \textit{Route} which is for the home page and it catches all unknown paths.

\subsection{MyAlgo}
\label{subsection:myalgo}
\begin{figure}[H]
\includegraphics[scale=0.55]{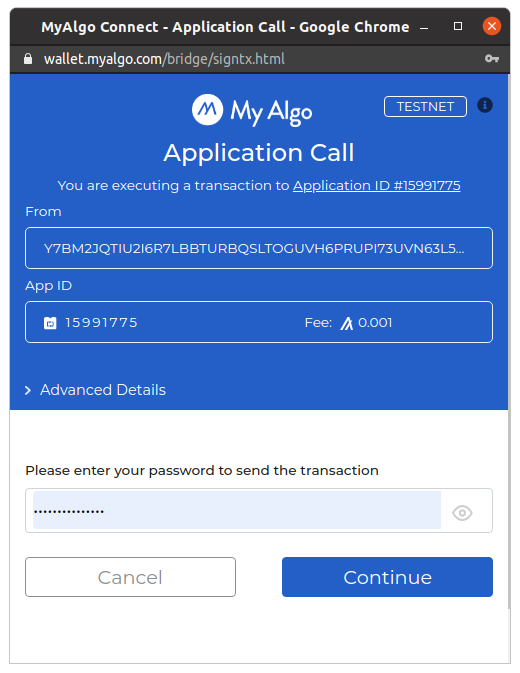}
\centering
\caption{Signing Transaction Using MyAlgo Connect}
\label{fig:myalgoconnect}
\end{figure}

To use the web application, a user must first create a MyAlgo Wallet account. They can then share their public Algorand addresses with the web application and make use of the MyAlgo Connect functionality to sign transactions securely, without having to expose their private keys. Figure \ref{fig:myalgoconnect} is a screenshot of a generated Algorand transaction from which the user can review and sign. 

Other than MyAlgo connect, the only other alternative for allowing users to sign transactions without having to store private keys is AlgoSigner \cite{algosigner}. AlgoSigner has the same functionality as MyAlgo Connect, however it is only available as a chrome extension so was not preferred. 

One limitation with MyAlgo Connect is that you cannot sign grouped atomic transactions, see section \ref{subsection:atomictransfers}. This provides a negative experience for the user as they have to sign each transaction separately and unless they look at the underlying code, they are unaware of any grouped atomic transactions. A malicious website could use this behaviour to steal funds from an account. For example, the user could sign a payment transaction expecting it to be grouped with a returned asset transfer. The website could then submit signed transaction alone.

Therefore the current implementation is not ideal and would not be suitable in a production environment. The MyAlgo team say that MyAlgo Connect will support grouped transactions in the next release so the problem should be resolved in the near future.

\subsection{Issuer}
\label{subsection:issuer}
To issue a bond the issuer needs to fill in a form with the bond parameters: number of bonds to mint, number of coupon payments, start buy date, end buy date, maturity date, bond cost, bond coupon value and bond principal amount. For the purpose of the demo and allowing users to try out all the available interactions, they also specify the Algorand address for the green verifier and financial regulator. In practise, for the green verifier and financial regulator fields, there would either be a drop down option where the issuer can select from a few choices or they would have no say in the matter.

When the issuer submits the form, the request is sent to the backend where the parameters are encoded into the smart contracts and are deployed. A new ASA is also created. The reason why these transactions are not created for the issuer to sign and submit is because of the current limitations in Algorand. Each address is only allowed to create up to 10 stateful smart contracts. There are two stateful smart contracts per issuance which would limit a single address to five green bond issuances in total. This can be handled by having the issuer generate and link new Algorand addresses but an easier solution is to just deploy the stateful smart contracts on behalf of the issuer. This is done in the backend by funding a new Algorand account which is used only ever once to both mint the green bond and create the applications. No additional keys have to be stored as the address will never need to be used again.

The page also lists all the green bonds the account has issued. From there an issuer can upload the use of proceeds and periodic reports, fund the stablecoin escrow account, as well as review the ratings they have received. To upload a document, an issuer first chooses a locally stored PDF which is uploaded to IPFS. They are then asked to sign a zero Algo payment transaction which contains the content identifier for the file, as explained in section \ref{subsection:ipfs}.

\subsection{Investor}
The investor can view for sale, ongoing and expired green bonds. These are each listed in their own data table from which they are able to filter by cost, expiry date etc. To obtain the green bond offerings, a request is sent to the NodeJS server which retrieves the bond and stateful smart contracts identification numbers from the PostgreSQL database. 

The MyAlgo developer API is then queried to obtain blockchain readings concerning the green bonds. The number of minted bonds, the balance in the stablecoin escrow account, the number of coupons paid, among other values, are all returned. 

The user can enter into a particular green bond by clicking on an entry in the table. From the new view, the user can register for the green bond which enables them to purchase the bond. Registering is defined as opting into the green bond ASA so that their Algorand address can hold them, and opting into one of the stateful smart contracts so that we can store information in their local state. All of this is hidden from the user in that they only need to click on the \textit{Register} button and these transactions are automatically generated for them to sign.

When purchasing the bond, the user can enter the number of bonds up to six decimal places (fractional asset ownership). Bond holders can then use the other exposed functionality to claim coupons, principal or default. The application will for example, verify they own some bonds, check that they have claimed all the coupons available and make sure their is insufficient funds in the stablecoin escrow account to pay all the remaining money owed, before allowing the investor to claim a default. If all these pass, then the investor can submit a default claim and the application will determine the amount of money owed to them which is embedded into the generated transactions. If the investor tries to get around the checks by creating these transactions themselves then the smart contract logic will reject them since they have their own similar checks.

Another part of the DApp is the ability for investors to trade their bonds in the secondary market. Green bond owners can set the number of bonds they are willing to sell and at which price. Other investors can then search and filter through all these trade offers and decide to accept them.

Investors can use the interface to generate the delegated logic signatures which is stored in the PostgreSQL database so that when someone wants to use a particular trade offer, the signature is retrieved for the purpose of signing some of the transactions in the atomic transfer. Section \ref{subsection:trade} describes the blockchain part of the implementation in more detail.

\subsection{Green Verifier}
The green verifier has a similar view to an issuer, except that instead of uploading PDF report and reviewing ratings, they review PDF reports and submit ratings. To do this, the green verifier first selects the green bond from a list of all the bonds they are a green verifier of.

The uploaded PDFs are retrieved by searching the blockchain transactions of the issuer for the content identifiers which is used as a lookup into IPFS. The interface tells the green verifier the period for which they can add a rating for as demonstrated in Figure \ref{fig:reportratingtimeline} and then they simply select the number of appropriate stars. A transaction is generated with these values for them to sign.  

\subsection{Financial Regulator}
The financial regulator can select from all the green bonds that they are the financial regulator for. After selecting a particular green bond, the financial regulator can see the list of all registered investors for the bond and their respective balances, all obtained from reading the blockchain. 

By default both the issuer and all investors must be approved by the financial regulator. The financial regulator simply clicks on the button with their desired action and the corresponding transaction is generated for them to sign.

\chapter{Evaluation}
\label{chapter:evaluation}


The evaluation of the project is determined by the extent to which the original objectives of the project were met, defined in section \ref{section:objectives}. The first objective can be evaluated by reviewing the functionality, security and cost of the blockchain implementation. The second objective can be evaluated by looking at the limitations of Algorand and areas which are lacking in its development.

\section{Algorand Fees}
There are no currently available tools that estimate the cost values for you so all the costs were manually calculated using the Algorand documentation as a reference. To review the fees in Algorand, please see section \ref{subsection:fees}.

The exact cost is dependant on a number of factors. For example an issuer has to submit an additional transaction for every report they upload. Therefore the final values will be expressions from which we can substitute example values into.

To issue the green bond there are fees to mint a new ASA, deploy the stateful smart contracts and supply the contract account with the minimum balance fee. The calculation can be seen in table \ref{tab:issuancecost}, where n is the number of uploaded reports and c is equal to:
\[ c = \left\lceil \frac{CouponRounds + 1}{8} \right\rceil \] 

\begingroup
\renewcommand{\arraystretch}{1.2}
\begin{table}[H]
    \centering
    \begin{adjustbox}{max width=\textwidth}
    \begin{tabular}{c|c|c|c|c}
    \textbf{Action}                                                                   & \textbf{\begin{tabular}[c]{@{}c@{}}Amount\\ (microAlgos)\end{tabular}} & \textbf{\begin{tabular}[c]{@{}c@{}}Minimum\\ Balance Fee\\ (microAlgos)\end{tabular}} & \textbf{\begin{tabular}[c]{@{}c@{}}Transaction\\ Fee\\ (microAlgos)\end{tabular}} & \textbf{\begin{tabular}[c]{@{}c@{}}Total\\ (microAlgos)\end{tabular}} \\ \hline
    Create new ASA                                                                    & 0                                                                      & 100,000                                                                               & 1,000                                                                             & 101,000                                                               \\ \hline
    \begin{tabular}[c]{@{}c@{}}Fund contract\\ accounts\end{tabular}                  & 202,000                                                                & 0                                                                                     & 1,000                                                                             & 203,000                                                               \\ \hline
    \begin{tabular}[c]{@{}c@{}}Send green bond to\\ escrow and configure\end{tabular} & 0                                                                      & 0                                                                                     & 2,000                                                                             & 2,000                                                                 \\ \hline
    Deploy Main App                                                                   & 0                                                                      & 184,000                                                                               & 1,000                                                                             & 185,000                                                               \\ \hline
    Deploy Manage App                                                                 & 0                                                                      & 100,000 + 50,000c                                                                     & 1,000                                                                             & 101,000 + 50,000c                                                     \\ \hline
    Update Apps                                                                       & 0                                                                      & 0                                                                                     & 2,000                                                                             & 2,000                                                                 \\ \hline
    Upload Report                                                                     & 0                                                                      & 0                                                                                     & 1000n                                                                             & 1000n                                                                 \\ \hline
                                                                                      &                                                                        &                                                                                       &                                                                                   & 594,000 + 50,000c + 1,000n                                           
    \end{tabular}
    \end{adjustbox}
    \caption{Issuance Cost}
    \label{tab:issuancecost}
\end{table}
\endgroup

The investor has a minimum balance to opt into receiving the green bond and storing local state in the main application. After that, the investor simply pays for transaction fees. The cost for each investor function is specified in table \ref{tab:investorcosts}.

\begingroup
\renewcommand{\arraystretch}{1.2}
\begin{table}[H]
    \centering
    \begin{tabular}{c|c|c|c|c}
    \textbf{Action} & \textbf{\begin{tabular}[c]{@{}c@{}}Amount\\ (microAlgos)\end{tabular}} & \textbf{\begin{tabular}[c]{@{}c@{}}Minimum\\ Balance Fee\\ (microAlgos)\end{tabular}} & \textbf{\begin{tabular}[c]{@{}c@{}}Transaction\\ Fee\\ (microAlgos)\end{tabular}} & \textbf{\begin{tabular}[c]{@{}c@{}}Total\\ (microAlgos)\end{tabular}} \\ \hline
    Opt into ASA    & 0                                                                      & 100,000                                                                               & 1,000                                                                             & 101,000                                                               \\ \hline
    Opt into App    & 0                                                                      & 184,000                                                                               & 1,000                                                                             & 185,000                                                               \\ \hline
    Buy             & 1,000                                                                  & 0                                                                                     & 3,000                                                                             & 4,000                                                                 \\ \hline
    Trade Sell      & 1,000                                                                  & 0                                                                                     & 3,000                                                                             & 4,000                                                                 \\ \hline
    Trade Buy       & 0                                                                      & 0                                                                                     & 1,000                                                                             & 1,000                                                                 \\ \hline
    Claim Coupon    & 1,000                                                                  & 0                                                                                     & 3,000                                                                             & 4,000                                                                 \\ \hline
    Claim Principal & 2,000                                                                  & 0                                                                                     & 4,000                                                                             & 6,000                                                                 \\ \hline
    Claim Default   & 2,000                                                                  & 0                                                                                     & 4,000                                                                             & 6,000                                                                
    \end{tabular}
    \caption{Investor Costs}
    \label{tab:investorcosts}
\end{table}
\endgroup

The green verifier and financial regulator pay in total 1,000 microAlgos per transaction. This would be to provide a green rating or to freeze the bond.

For a green bond that has annual coupon payments for 10 years, assuming one report per period and ignoring trading (which has a very small fee), we have the following costs:
\begin{itemize}
    \item \textbf{Issuer} 1.104 Algos or \$1.13
    \item \textbf{Investor} 0.336 Algos or \$0.34
    \item \textbf{Green Verifier} 0.011 Algos or \$0.01
    \item \textbf{Financial Regulator} Negligible 
\end{itemize}

These costs are minimal; if Algorand's market cap reached Ethereum's market cap, the cost of issuance would be \$103.60 and the investor would pay \$30.84 in total.

\section{Green Rating}
As we have seen, the green verifier submits green ratings for a bond before each coupon. For every rating dropped from 5 stars, the coupon payment values increase by 10\% compounded, i.e. if the coupon rate is 2\% (for 5 stars) then the coupon rate for 3 stars is 2.42\%. Section \ref{section:greenrating} covers this in more details.

\begin{figure}[H]
\includegraphics[scale=0.35]{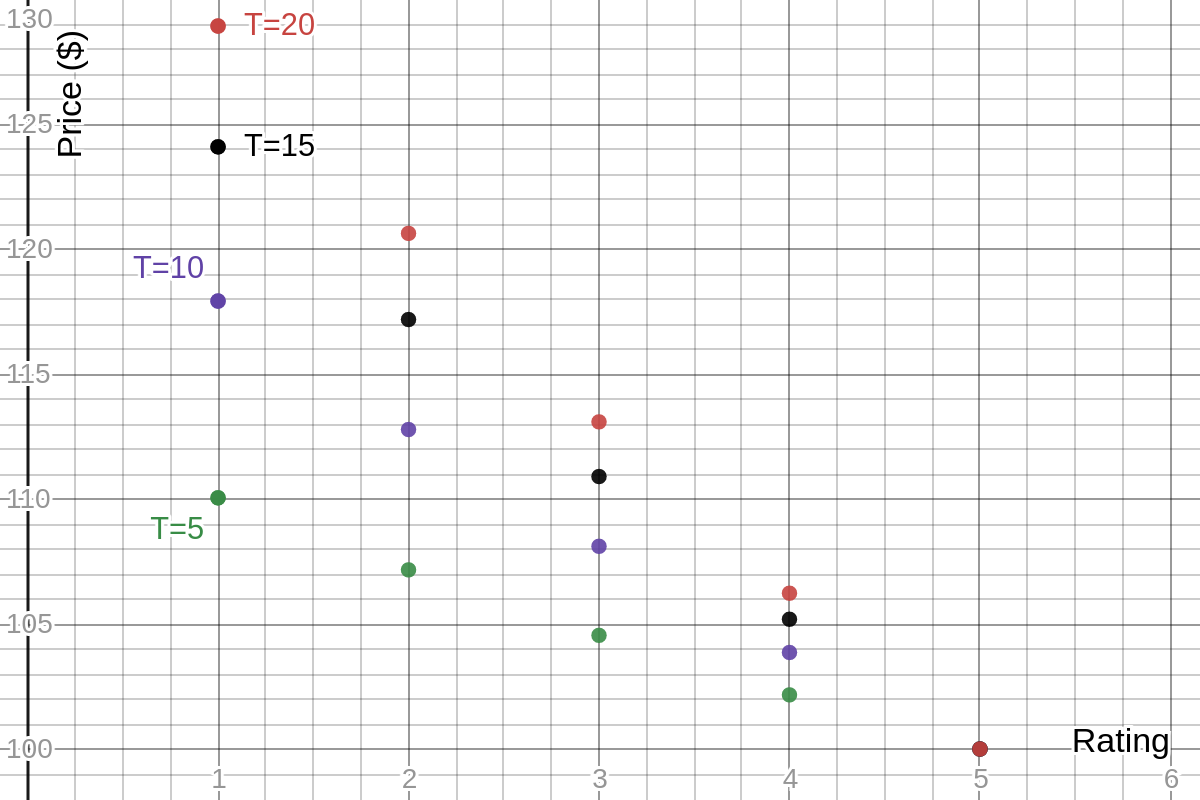}
\centering
\caption{Green bond price against green rating for various time periods}
\label{fig:bondpriceontimeperiod}
\end{figure}

\begin{figure}[H]
\includegraphics[scale=0.35]{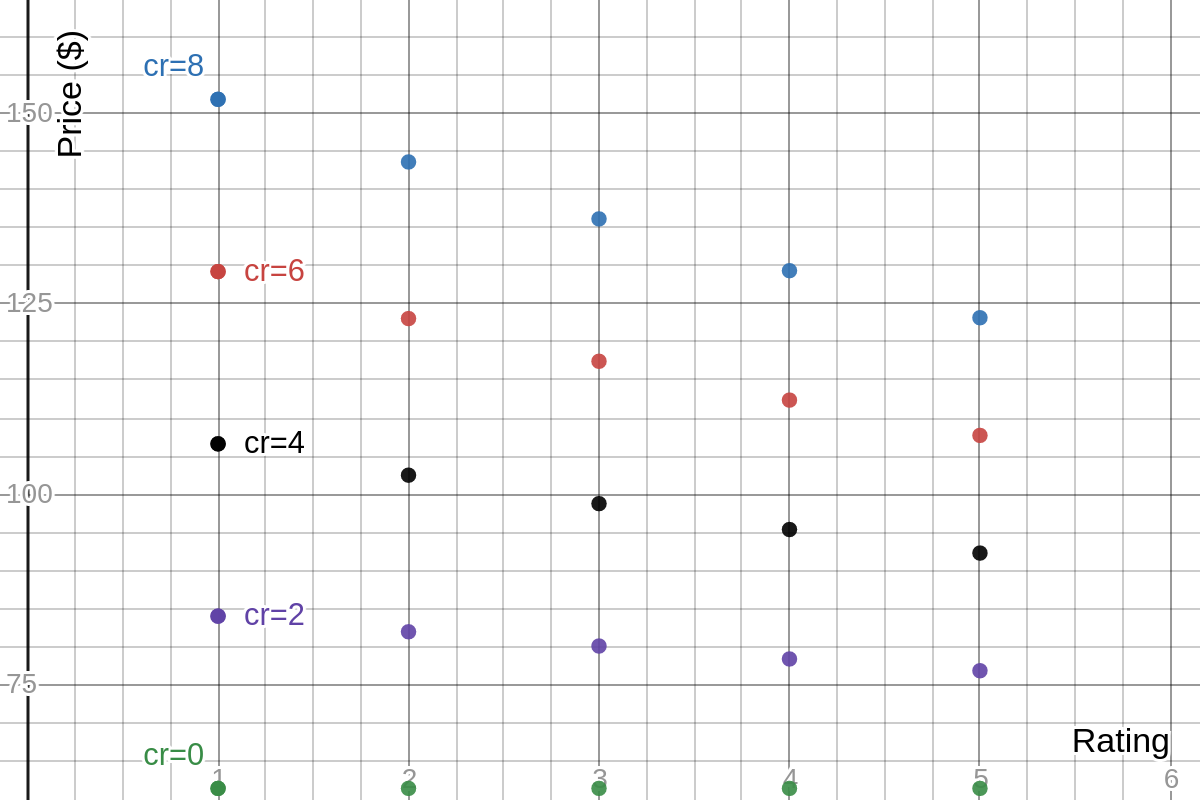}
\centering
\caption{Green bond price against green rating for various coupon rates}
\label{fig:bondpriceoncouponrate}
\end{figure}

We can use the bond price formula in section \ref{section:bonds} to evaluate the effect the green ratings have on the theoretical price of the bond. Figures \ref{fig:bondpriceontimeperiod} and \ref{fig:bondpriceoncouponrate} plots the price of the green bond against the green rating. We assume that the green bond received the same green ratings for all its coupons.

For figure \ref{fig:bondpriceontimeperiod}: the bond pays an annual coupon rate of 5\% (for 5 stars), has a face value of \$100 and has a discount rate of 5\%. Each colour represents a different length of time until maturity, and these range from 5 years to 20 years. 

The graph shows that bad green performance over increased time periods results in larger price increases. Investors receive higher coupon payments so the price of the bond is driven up. We can view green ratings similar to credit ratings where even though investors may get a higher return on their investments, there is an added risk where their funds may not be used for suitably green projects.

For figure \ref{fig:bondpriceoncouponrate}: the bond has a face value of \$100, a discount rate of 5\% and has 10 years until maturity. Each colour represents a different annual coupon rate (for 5 stars), and these range from 0\% (zero coupon bond) to 8\%.

Green ratings have no effect on the price of a zero coupon bond as the ratings are tied to the coupon payments. This reflects are architecture where we expect the issuer to supply data and documentation at every coupon period regarding the green status of their project. If there are no coupons then there is no reporting (other than the use of proceeds). 

Overall we can see that as the coupon becomes less of a factor in the pricing of the bond, the green ratings have a smaller impact on the price as well. We could look into potentially reducing the penalties as the coupon rate in relation to the face value increases, and introducing a penalty for the principal which decreases as the face value in relation to the coupon rate decreases.

\section{Security}
The Algorand recommend guidelines were followed when writing the smart contracts. The contracts verify that there in never any rekeying which prevents a malicious actor from being able to take control of an Algorand account through an unchecked transaction. There are also checks that the close-to property of a transaction is not set so funds cannot be stolen from the escrow accounts. 

Unit tests were written to verify the behaviour of the smart contracts and possible edge cases. These ensure that certain transactions fail and that malformed arguments are handled appropriately. 

The stateful smart contracts also disables the update functionality. A compromised issuer account would not have any impact on the security of their issued green bonds. Once the green bond is created and the funds are received, the issuer no longer interacts with the applications.

Two concerns though are the green verifier and financial regulator. Through their rating, the green verifier directly impacts the money the issuer owes. The financial regulator has the ability to freeze the bond entirely. Both of these actors are assumed to be trustworthy but there should be additional safeguards in place to protect against malicious behaviour or leaked private keys. 


\section{Contributions to The Algorand Ecosystem}
Since Algorand is a relatively new technology, the ecosystem as a whole is underdeveloped. When working with community tools many bugs were found which stunted project progress. Some of the most important ones were:

\begin{itemize}
    \item Vulnerability in MyAlgo Wallet where you can spam an account with millions of assets. The attack goes as follows: let's say we have two new Algorand accounts, one controlled by Alice and the other controlled by Bob. Alice then sends Bob a zero amount of an ASA. The transaction fails on the blockchain since Bob never opted into the asset, however the MyAlgo displays the ASA in Bob's list of holdings. Alice can repeatedly do this at no cost since the transactions would never be approved.  
    \item Major bug in Algo Builder \cite{algobuildertxnbug}, the only testing framework for Algorand smart contracts. The \textit{Txn} opcode always referenced the first transaction in a transaction group as opposed to the current transaction.
    \item Bug in Algo Builder \cite{algobuildercloseassetbug} where you could opt out of an asset using a clawback transaction. This behaviour is not allowed on the Algorand blockchain.
\end{itemize} 

As well as the above issues, an additional three issues were filed in the Algo Builder repositories detailing improvements that could be made that resulted in pull requests. There were also two minor bugs reported on the MyAlgo Connect signing tool that were also fixed.

\section{Algorand}
There are many challenges to working with Algorand. As outlined above, there are few available tools for developers to work with and some of them have not been thoroughly tested. 

As discussed in \ref{subsection:myalgo}, there exists two signing tools that allow the user to sign transactions using their locally stored private keys. One is AlgoSigner but that can only be used as a chrome extension. Therefore the project uses My Algo Connect. My Algo Connect has a severe limitation in that it does not support grouped transactions (AlgoSigner does not support them either). This means that a user has to sign each transaction separately without having the guarantee that it is being grouped with another transaction. For Algorand to be a viable option, grouped transaction signing must be added. After reaching out to the MyAlgo team about this, they responded that they plan on adding this functionality in their next major release.

There is also only one testing framework for Algorand smart contracts. At the moment, most developers are manually testing their smart contracts by using a private network and writing scripts. This methodology was used at the start of the project but became very tedious as the set up times are long and it is difficult to test smart contracts manually that are dependent on time.

To solve this issue, the project made use of the Algo Builder tool. The project became the first open source project using Algo Builder. Algo Builder was difficult to work with at first as it lacked documentation and had a number of bugs. In late May 2021, version 1.0 of Algo Builder was released which included new documentation and fixed many of the reported bugs. Algo Builder still remains limited in that it only supports up to TEAL2. All TEAL3 code has to be commented out so portions of the smart contracts must be tested solely manually.

There are also a number of problems with TEAL. Currently TEAL does not support looping which restricts what you can write. In addition there are size limitations on these contracts, see section \ref{subsection:teal}, which were quickly exceeded. To combat this, argument names had to be shortened and the stateful smart contract code had to be split into two stateful smart contracts that referenced each other. 

The supported languages for smart contract development are very basic. In the case of PyTeal, it is a simple wrapper of the TEAL language in which you still have to directly interact with the stack. In addition, all debugging must be done by first compiling to TEAL and then stepping through the code.

In May 2021, the Reach language \cite{reach}, which targets the Algorand and Ethereum blockchain, was released on MainNet. Reach uses a subset of JavaScript and is much more expressive than TEAL. It has a compiler which is integrated with a satisfiability-modulo-theories (SMT) theorem prover which verifies the correctness of the application. Since Reach came towards the end of the project, there was not enough time to test it out, however it looks like a great step to lowering the barrier to entry for Algorand.

\section{Comparison With Existing Solutions}
There are few comparable solutions for green bond issuance on the blockchain. Some companies are working on blockchain bonds more generally, but they are still in development. This presents a unique opportunity in the market for a platform like the one developed in this project.

\subsection{Green Assets Wallet}
One platform that is aimed towards green bonds is Green Assets Wallet (GAW) \cite{greenassetswallet}. GAW support green impact reporting using the Chromia blockchain. It uses the concept of \textit{Validators}, whose role it is is to assess the documentation and data provided by an issuer. This is similar to the green verifier in our case. However the platform is limited; it only allows investors to view this information but all bond transfers and payments are done off-chain. Our proposed implementation supports the green bond in all aspects, from issuance to  maturity, and thus benefits from immediate settlement and reduced transaction costs.

\chapter{Legal and Ethical Issues}
\label{chapter:ethics}

One area to consider are the inconsistencies between blockchain and the European Union's General Data Protection Regulation (GDPR) \cite{gdpr}. Article 17, \textit{right to be forgotten}, states that a data subject can withdraw consent and the controller has an obligation to erase the personal data. Everything written to Algorand is permanent so one has to be careful when handling personal data. For our use case, there are privacy concerns where potentially the green bonds one owns can help identify the individual that controls that Algorand account.

Green bond issuers also have Know Your Customer (KYC) and Anti-Money Laundering (AML) obligations. For this reason, blockchain bonds have in the past mainly been issued on permissioned blockchains with access to verified investors. In our smart contract implementation, the financial regulator must approve the green bond issuance and each account before they can purchase the green bond. This allows for off-chain KYC and AML checks. In addition, the financial regulator has the ability to freeze one's green bond assets.

There are a number of risks in issuing green bonds using blockchain. To combat cryptocurrency price volatility, we used stablecoins however they have their own downsides. Stablecoins are vulnerable to a bank or other financial institution's risks (see section \ref{subsection:stablecoin} for more details). There are also risks associated with the security of the underlying blockchain used. 

Smart contracts often cannot be altered once written to the blockchain. Thorough analysis must be done to ensure it is impossible for an attacker to exploit a vulnerability in the program. Blockchain green bonds will also lead to fewer intermediaries, potentially removing important checks and balances \cite{hsbc}.

All code developed and any associated intellectual property are licensed under the MIT license. This means that the project can benefit the wider community and help contribute to the green bond field. One possible unintended consequence is an increase in \textit{greenwashing}. Regulatory frameworks should be established to ensure green bond proceeds are spent appropriately.
\chapter{Conclusion}
\label{chapter:conclusion}

\section{Reflection}
The green bond market remains relatively small in comparison to the global bond market. The main obstacles to growth is: the perception of higher issuance costs, the risks of greenwashing for all stakeholders and lack of standardisation  \cite{greenbondmarket}. 

The work in this project demonstrates that blockchain can be used to substantially lower the costs and also provide transparency and accountability with regards to greenwashing. The settlement period can be entirely eliminated through disintermediation. We can also use a financial regulator to approve issuers and investors, thus complying with KYC / AML legal requirements.

The Algorand blockchain has many features that make it a great fit for green bond issuance. It has high throughput, low costs, and layer-1 support for assets and smart contracts. Its consensus mechanism also uses a lot less energy compared to other blockchains. However further development is needed around the Algorand ecosystem before it should be used in a production environment.

\section{Summary of Achievements}
We have developed a proof of concept application for green bond issuance on blockchain that is cost effective. The implementation supports the life cycle of green bonds from issuance to maturity. Fractional asset ownership and low issuance fees opens the market to smaller investors and issuers. Issuers can receive stablecoin from the proceeding of the bond and investors can receive stablecoin through the coupons and principal. The green bonds can also be traded in the secondary market at virtually no cost.

We created a novel system whereby an issuer can upload reports which are validated by a green verifier. The green ratings submitted has a direct impact on the issuer and the coupon amounts. The issuer not only suffers reputations damage, they are also hit with economic penalties.

Lastly the financial regulator is able to freeze tokens to ensure the integrity of the market. An investor must first be preapproved  before they are able to purchase the green bond, whether on the primary or secondary market. The blockchain also can be used to monitor the financial transactions of the green bond. Mechanisms are in place for investors to recover any remaining funds in the case of bond default.

\section{Future Work}
We believe that there are a number of opportunities for extending the project. Below are a selection of the most important areas for future work.

\subsection{Blockchain Oracles For Impact Reporting}
At the moment the issuer uploads reports detailing the green impact of their project. The green verifier must review this documentation and submit an appropriate green rating. Credibility can be improved by shifting the validation from the green verifier to tokenised data \cite{hsbc}. The first stage of this is integrating blockchain oracles that provide trusted off-chain data to the smart contracts. 

\subsection{Green Penalties}
The current formula for green penalties results in harsher penalties for green bonds with higher coupon rates. The most extreme case being zero coupon bonds, where the ratings have no impact. We can refine the formula to take into account the principal, and potentially scale the coupon and principal penalties according to their respective contribution to the price of the bond. One way we can do this for the coupon is by basing the penalty on the spread, the interest rate in excess of the risk free rate.

\subsection{Primary Market}
Currently the issuer sets a price up front. An investor must either pay that price initially, or buy the bond from another investor. However to better reflect the primary market experience, the issuer should be able to select if they would like to issue the green bond using a \textit{bought deal} or \textit{auction}.

\subsection{Automated Transactions}
In Algorand, smart contracts cannot create transactions, only approve them. Therefore an Algorand account needs to submit the appropriate transactions to for example claim the coupon, and the smart contracts verify that the amount of money being claimed is correct and so on. Up to now each investor has been responsible for claiming their own coupons but it would be preferable if these were claimed on their behalf.

A scheduling service can be deployed that does exactly this. An investor would sign logic detailing the transactions needed. A script would record the dates of the coupon payments and at each event, it would use the delegated signature to submit the transactions as if it came from the investor. A similar approach can be used in other areas including bond defaults and principal payments.

\appendix
\chapter{User Guide}
\label{userguide}

This section will be a resource for how to get setup with the application and interact with it. You can access the site using the URL \url{https://blockchain-bonds.herokuapp.com/}. Note that it may take some time for the page to initially load the website if there has not been any recent traffic. Similarly after logging in, it make take some time to retrieve the associated Algorand addresses linked to your account while the server is booting up.

The recommended browser to use when accessing the website is Google Chrome. Other browsers have not been tested although they should still all work.

The web application is designed so that you can try out all the different roles. You can be an issuer, investor, green verifier or financial regulator.

\section{Create An Account}

\begin{enumerate}
    \item Create an account by clicking on the \textit{Sign Up} button in the top right corner of the page. 
    \item Sign up by typing in your email and a password, or by using Google account.
\end{enumerate}

After logging in, you will be directed to the dashboard page. The dashboard page shows the selected Algorand address, as well as its Algo and stablecoin balances.

\begin{figure}[H]
\includegraphics[scale=0.47]{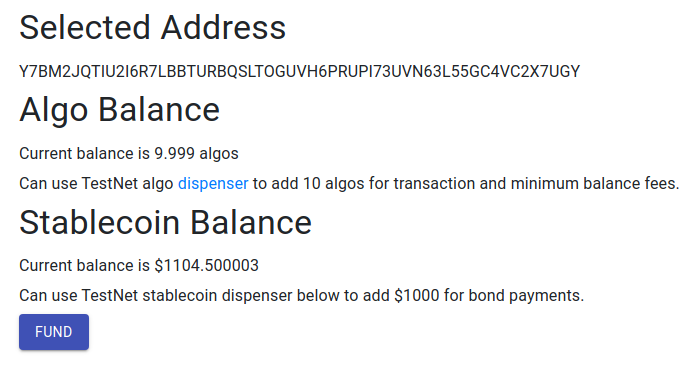}
\centering
\caption{Screenshot From Dashboard}
\label{fig:dashboard}
\end{figure}

\section{Set Up My Algo Wallet}

If this is your first time logging in, you will have to set up your account by connecting your Algorand addresses stored in your MyAlgo Wallet.

\begin{enumerate}
    \item Click on \textit{Settings} on the top right, next to \textit{Log Out}.
    
    \item Click on the \textit{CONNECT} button, a popup should appear. If you already have a MyAlgo account then you can skip to step 7.
    
    \begin{figure}[H]
    \includegraphics[scale=0.35]{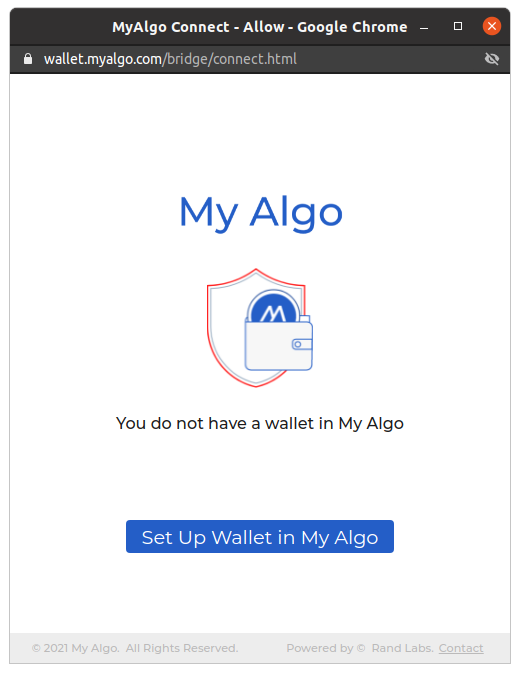}
    \centering
    \caption{MyAlgo Set Up Wallet Pop Up}
    \label{fig:myalgosetupwalletpopup}
    \end{figure}
    
    \item Click \textit{Set Up Wallet in My Algo}, a new tab for \url{https://wallet.myalgo.com/} should be opened.
    
    \item Click \textit{Access Now}, accept the terms of service and type in a password. The MyAlgo Wallet will only be accessible in that browser as all the keys are privately stored and thus not accessible through the Internet.
    
    \item On the top right, click on the drop down menu and select \textit{TESTNET}. 
    
    \begin{figure}[H]
    \includegraphics[scale=0.6]{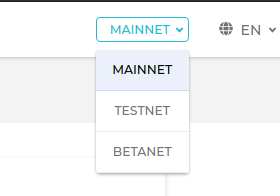}
    \centering
    \caption{Switch to TestNet in MyAlgo}
    \label{fig:myalgotestnet}
    \end{figure}
    
    \item Select the second option from the top \textit{New Wallet}. Follow the instructions to generate a new Algorand address. You can name the wallet whatever you like.
    
    \item Return to the settings page and click \textit{CONNECT} again. This time a pop should appear asking you to type in your MyAlgo password. 
    
    \begin{figure}[H]
    \includegraphics[scale=0.4]{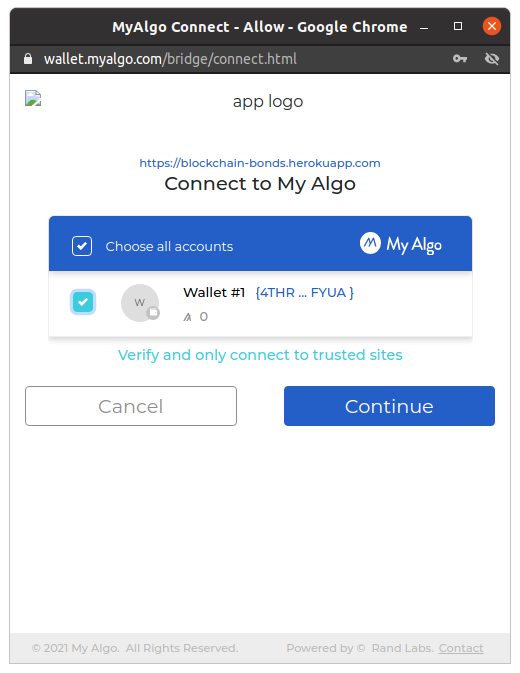}
    \centering
    \caption{MyAlgo Connect Addresses Pop Up}
    \label{fig:myalgoconnectaddress}
    \end{figure}
    
    \item Tick the addresses you would like like to connect to and press \textit{Continue}. The addresses should now be listed under \textit{Connected Account}. From here you can select which address you would like to interact with the blockchain from.
    
    \begin{figure}[H]
    \includegraphics[scale=0.5]{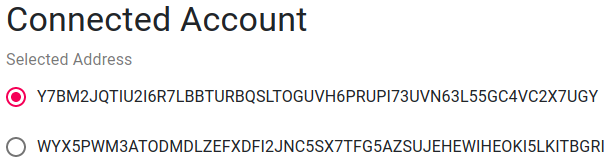}
    \centering
    \caption{MyAlgo Connect Addresses Pop Up}
    \label{fig:connectedaddresses}
    \end{figure}
\end{enumerate}

\section{Fund Algos}
\label{section:fundalgos}
Before you can interact with the Algorand blockchain, you first need to fund your account with Algos.

\begin{enumerate}
    \item Click on \textit{Dashboard} on the top left of the page.
    \item Click on \textit{dispenser} under the \textit{Algo Balance} heading. This should open a new tab.
    \item Fill in the form with the address you would like to add Algos to and click dispense. If you return to the dashboard page, you should see the updated Algo balance. 10 Algo should be plenty.
\end{enumerate}

\section{Fund Stablecoin}
Before you can purchase a bond / fund an escrow account as an issuer, you need to have some stablecoin in your Algorand account. 

\begin{enumerate}
    \item Click on \textit{Dashboard} on the top left of the page.
    \item Click on \textit{FUND} under the \textit{Stablecoin Balance} heading. You should see the updated stablecoin balance.
    
    \begin{figure}[H]
    \includegraphics[scale=0.6]{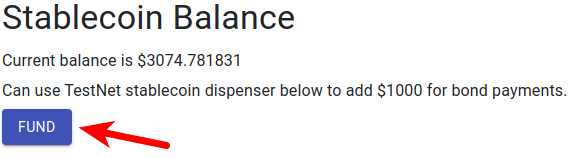}
    \centering
    \caption{Issue Bond Form}
    \label{fig:fundstablecoin}
    \end{figure}
\end{enumerate}

\section{Issuer}
To take the role of an issuer, click on \textit{Issuer} in the top navigation bar. The issuer has a number of functionalities: they can issue green bonds and manage their existing ones. This includes funding the account from which the coupon and principal payments will be taken from and uploading their green documentation. 

To issue a new bond:

\begin{enumerate}
    \item Click on \textit{ISSUE NEW BOND} at the bottom of the page.
    \item Fill in the form with the green bond parameters. If you would like to be the green verifier and financial regulator then you can give an Algorand account address that you control in these fields. In practise you would not have the ability to set the green verifier and financial regulator.
    
    \begin{figure}[H]
    \includegraphics[scale=0.55]{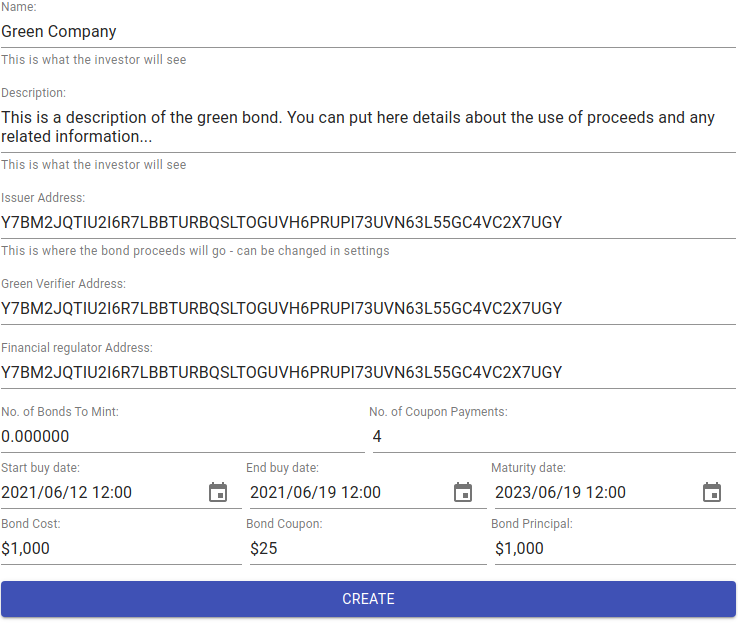}
    \centering
    \caption{Issue Bond Form}
    \label{fig:issuebondform}
    \end{figure}
    
    \item Click \textit{CREATE}. This will take a few minutes to set up in the background. After everything has been created, you should be able to see the bond listed in your issuer page.
    \item Before investors can purchase the new green bond, it must be first approved by the financial regulator by them unfreezing the bond, see section \ref{section:financialregulator} for how to do this.
\end{enumerate}

To upload reports for a green bond you have issued:

\begin{enumerate}
    \item Click on the green bond you would like to upload a report on from the table. 
    \item If at the appropriate time (before the start buy date for the use of proceeds or in a coupon period for a report) then there will be a button enabled to upload a PDF. Click on \textit{Upload PDF For <X>} and choose a PDF from your local file system.
    \item Sign the generated transaction. You should see the uploaded report after refreshing the page.
\end{enumerate}

To fund the escrow account of a green bond you have issued:

\begin{enumerate}
    \item Click on the green bond you would like to upload a report on from the table. 
    \item At the bottom of the page, enter the dollar amount you would like to transfer into the escrow account, and click the button to the right.
    \item Sign the generated transaction. You should see the uploaded report after refreshing the page.
\end{enumerate}

\section{Investor}
To take the role of an investor, click on \textit{Investor} in the top navigation bar. You will have a number of options to select from depending on what you would like to do.


To buy a bond in the primary market:

\begin{enumerate}
    \item Click on \textit{Bonds For Sale} and select the bond you would like to purchase.
    \item You can review information about the bond, like its timeline, uploaded reports and green ratings.
    \item Before you can own the green bond you must first register by clicking on \textit{REGISTER}. You will then have to wait until your account has been approved by the financial regulator.
    \item After you have been approved, under the \textit{Purchase} heading, specify the number of bonds you would like and click the buy button. 
    \item Sign the generated transactions. You should see that the number of bonds you own update.
\end{enumerate}

The buttons for claiming a coupon, principal or default are enabled when appropriate. If disabled, you can hover over the button to see the reason why.

To sell a bond in the secondary market:

\begin{enumerate}
    \item Navigate to the green bond page for the one you would like to trade.
    \item Under the \textit{Trade} heading, specify how many bonds you would like to trade and click the button to the right. \item Sign the generated transactions. You should see the number of bonds that can be traded update.
    \item Generate a trade logic signature by filling in the fields below, specifying the price per bond and the expiry date of the trade offer and click the button to the right.
    \item Sign the generated transactions.
    \item If you return to the investor page and click \textit{My Trades}, you should see your newly created trades there.
\end{enumerate}

To buy a bond in the secondary market:

\begin{enumerate}
    \item Click on \textit{Live Trade Offers} and select the trade offer you would like to take up.
    \item If you have not already registered for the bond, you must first do so by clicking on \textit{REGISTER}. You will then have to wait until your account has been approved by the financial regulator.
    \item Specify the number of bonds you would like to buy, up to the maximum number available, and click on the button to the right.
    \item Sign the generated transactions. You should see that the number of bonds you own update.
\end{enumerate}

\section{Green Verifier}
To take the role of a green verifier, click on \textit{Green Verifier} in the top navigation bar. The page will list all the green bonds that you are a green verifier for. To add a green rating for a bond:

\begin{enumerate}
    \item Click on the green bond you would like to upload a rating for from the table. 
    \item You can review the issuer's uploaded PDFs by opening a time period and clicking on the links.
    \item If at the appropriate time (before the start buy date for the use of proceeds or in a coupon period for a report) then there will be a button enabled to submit a rating. Specify the number of stars to give and click on \textit{Add Rating For <X>}.
    \item Sign the generated transaction. You should see the page update with the submitted rating.
\end{enumerate}

\section{Financial Regulator}
\label{section:financialregulator}
To take the role of an financial regulator, click on \textit{Financial Regulator} in the top navigation bar. From there you can see all the accounts that have registered for the green bond.

To freeze / unfreeze the green bond as a whole then click on the button at the top of the page and sign the generated transaction. When a green bond is first issued, it is created frozen so unfreezing it is the financial regulator approving the bond for sale.

To freeze / unfreeze the green bond for a specific account, click on the switch next to the account you would like to perform the action to. When an account registers for a green bond, their account is frozen so unfreezing it is the financial regulator approving that account to own, claim money from and trade the green bond.

\bibliographystyle{unsrtnat}
\bibliography{bibs/bibliography}

\end{document}